\documentclass[10pt,journal,compsoc]{IEEEtran}
\usepackage{amsmath}
\usepackage{cite,dsfont}
\usepackage{graphicx}
\usepackage{amsfonts}
\usepackage{latexsym,dsfont}
\usepackage{booktabs}
\usepackage{algorithm}
\usepackage{algorithmic}
\usepackage{color,soul}
\usepackage{times,booktabs}
\usepackage{epsfig, graphics}
\usepackage{bm}
\usepackage{subcaption}
\usepackage{graphicx,wrapfig,lipsum}
\usepackage{array}
\usepackage{graphicx}
\usepackage{multirow}
\usepackage{enumerate}

\newcommand\MyBox[2]{
	\fbox{\lower0.75cm
		\vbox to 1.7cm{\vfil
			\hbox to 1.7cm{\hfil\parbox{1.4cm}{#1\\#2}\hfil}
			\vfil}%
	}%
}

\begin{document}
	\renewcommand{\figurename}{Figure}
	\newtheorem{theorem}{Theorem}	
	\newtheorem{lemma}{Lemma}
	\newtheorem{conjecture}{Conjecture}
	\newtheorem{corollary}{Corollary}
	\newtheorem{definition}{Definition}
	\newtheorem{scheme}{Scheme}
	\newcommand{\argmax}{\arg\!\max}
	
	\newcommand{\rev}[1]{{\color{black}#1}} 
	\newcommand{\reev}[1]{{\color{black}#1}}
	\newcommand{\rav}[1]{{\color{black}#1}}

	\newcommand{\pound}{\operatornamewithlimits{\gtrless}}
	\IEEEoverridecommandlockouts
	
	\title{DeepWiFi: Cognitive WiFi with Deep Learning 
		\thanks{K. Davaslioglu, S. Soltani, and Y. E. Sagduyu are with Intelligent Automation, Inc. Rockville, MD 20855, USA. Email:\{kdavaslioglu, ssoltani, ysagduyu\}@i-a-i.com. T. Erpek is with Intelligent Automation, Inc. and Virginia Tech., Electrical and Computer Engineering, Arlington, VA, USA; Email: terpek@vt.edu.} \thanks{DISTRIBUTION STATEMENT A. Approved for public release. Distribution is unlimited.}
		\thanks{\textsuperscript{\textcopyright} 2019 IEEE. Personal use of this material is permitted. Permission from IEEE must be
			obtained for all other uses, in any current or future media, including 
			reprinting/republishing this material for advertising or promotional purposes, creating new collective works, for resale or redistribution to servers or lists, or reuse of any copyrighted component of this work in other works.
		}
		\author{\IEEEauthorblockN{Kemal Davaslioglu, Sohraab Soltani, Tugba Erpek, and Yalin E. Sagduyu \vspace{-0.2in}}
		}
	}
	\maketitle
	
	\begin{abstract}
		We present the DeepWiFi protocol, which hardens the baseline WiFi (IEEE 802.11ac) with deep learning and sustains high throughput by mitigating out-of-network interference. DeepWiFi is interoperable with baseline WiFi and builds upon the existing WiFi's PHY transceiver chain without changing the MAC frame format. Users run DeepWiFi for i) RF front end processing; ii) spectrum sensing and signal classification; iii) signal authentication;  iv) channel selection and access; v) power control; vi) modulation and coding scheme (MCS) adaptation; and vii) routing. DeepWiFi mitigates the effects of probabilistic, sensing-based, \reev{and adaptive jammers}. RF front end processing applies a deep learning-based autoencoder to extract spectrum-representative features. Then a deep neural network is trained to classify waveforms reliably as idle, WiFi, or jammer. \reev{Utilizing channel labels, users effectively access idle or jammed channels, while avoiding interference with legitimate WiFi transmissions (authenticated by machine learning-based RF fingerprinting) resulting in higher throughput.} Users optimize their transmit power for low probability of intercept/detection and their MCS to maximize link rates used by backpressure algorithm for routing. \reev{Supported by embedded platform implementation}, DeepWiFi provides major throughput gains compared to baseline WiFi and another jamming-resistant protocol, especially when channels are likely to be jammed and the signal-to-interference-plus-noise-ratio is low.
	\end{abstract}
	
	\section{Introduction}
	Cognitive radios provide wireless communication systems with the capability to perceive, learn and adapt to spectrum dynamics. The existing communication systems such as WiFi can greatly benefit from design concepts of cognitive radio that cover various detection, classification, and prediction tasks. These cognitive radio tasks can be potentially performed by \emph{machine learning} that adapt with spectrum data without being explicitly programmed \cite{Clancy2007, Thilina2013, Alsheikh2014, Chen2017}. In particular, \emph{deep neural networks} have strong potential to process and analyze rich spectrum data. Examples of deep learning applications for cognitive radio tasks include modulation classification with convolutional neural networks (CNNs) \cite{OShea2016}, spectrum sensing with CNNs \cite{Lee2017} and generative adversarial networks (GANs) \cite{Kemal2018}, and anti-jamming and power control with feedforward neural networks (FNNs) \cite{Yi2018,Terpek18}. 
	
	\reev{In this paper, we present the DeepWiFi protocol that aims to harden the baseline IEEE 802.11ac WiFi with the application of machine learning (in particular, deep learning) to improve the throughput performance and enhance its security in the presence of jammers causing \emph{out-of-network interference}. We consider three types of jammers that are modeled as probabilistic, sensing-based, or adaptive jammers}. \rev{DeepWiFi leverages the advances in machine learning algorithms supported by emerging computational capabilities. The scope of DeepWiFi is to show how these algorithms can be used to enhance the performance and security in multi-hop mobile ad hoc networks (MANETs).
		Potential application areas of DeepWiFi include internet-of-things (IoT), sensors, and first responder communications.} One key aspect of DeepWiFi is to support \emph{interoperability} with baseline WiFi while refraining from any detrimental effect on baseline WiFi operations. For that purpose, DeepWiFi uses the PHY transceiver chain of baseline WiFi and does not change its medium access control (MAC) frame format. The DeepWiFi protocol stack involves the physical (PHY), link/MAC, and network layer algorithms that operate in a distributed manner. \rav{DeepWiFi supports multi-hop routing compared to the standard use of IEEE 802.11ac that relies on access points to connect individual nodes. As an extension, WiFi Direct can be utilized to support multi-hop communications \cite{Funai17} and the IEEE 802.11s amendment is specifically designed for mesh networking that incorporates multi-hop communications.} While this paper presents our findings using a WiFi setting, algorithms  of DeepWiFi can be extended to other frequency bands and waveforms as well.
	
	\subsection{Summary of Contributions}
	DeepWiFi consists of seven steps: 1) RF front end processing; 2) spectrum sensing and signal classification; 3) signal authentication;  4) channel selection and access; 5) power control for Low Probability of Intercept and low probability of detection (LPI/LPD); 6) modulation and coding scheme (MCS) adaptation; and 7) routing. \reev{The main contributions of this paper are the algorithms in steps 1-4, namely RF front end processing; spectrum sensing and signal classification; signal authentication; and channel selection and access. Steps 5-7 on power control, adaptive modulation and coding,  and routing are used in other similar protocols and included to provide the remaining functionalities (preserving the layered architecture of the protocol stack), implement and assess the full protocol stack, and compare its performance with other protocols.} While executing these steps, DeepWiFi controls the selection of transmit power, MCS ID, authenticated signal, channel ID, neighbor ID and flow ID, and provides the selections to the IEEE 802.11ac transceiver chain. 
	
	There are three tasks that are enhanced by the use of \reev{machine learning} in the DeepWiFi protocol. 
	\begin{enumerate}  
		\item \emph{RF front end processing}: \reev{Each user applies \emph{autoencoding} to the in-phase and quadrature (I/Q) components of the data collected on sensed channels and  extracts spectrum features.} The I/Q data can be collected by using additional receivers or extracted by using firmware dispatches on existing WiFi chips such as discussed in \cite{nexmon}. We show that autoencoding achieves low reconstruction loss (0.2\%) in dimensionality reduction compared to other methods such as Principal Component Analysis (PCA) and t-distributed Stochastic Neighbor Embedding (t-SNE). The outputs of the RF front end processing are input to the signal classification module.  
		\item \emph{Spectrum sensing and signal classification}: Each user applies \emph{deep learning} (FNN or CNN) to the features obtained by RF front end processing  in order to classify each channel as idle (I), jammed (J) or used by another WiFi device (W). We show that both FNN and CNN achieve over 98\% accuracy, whereas the accuracy of Support Vector Machine (SVM) is limited to 66\%. 
		\item \emph{Signal authentication}: Each user applies machine learning-based RF fingerprinting to authenticate signals. Radio hardware effects on RF signals are used as features to detect outliers. We show that the detection accuracy by Minimum Covariance Determinant (MCD) is close to 90\%, whereas SVM and Isolation Forest can only achieve around 69\% and 70\% accuracy, respectively. This signal authentication capability of DeepWiFi protects WiFi against replay attacks.  
		\item \reev{\emph{Channel access}: As opposed to baseline WiFi (where a user backs off regardless of the type of interference), each user in DeepWiFi runs the signal classification,  backs off when the interference is from another user participating in DeepWiFi protocol, or does not back off when the interference is from a jammer. This way, DeepWiFi users continue communications (in a degraded mode) and still achieve a non-zero throughput (instead of backing off) through adaptation of power, modulation, and coding scheme.}
	\end{enumerate}
	
	In this paper, a distributed network of users running DeepWiFi is simulated in the presence of probabilistic and sensing-based jammers. MATLAB WLAN Toolbox is used to generate realistic WiFi signals and channels. Simulation results show that as the jamming effect (either intentional adversarial jamming or in the form of out-of-network interference) increases, the throughput of baseline WiFi drops quickly to zero, whereas DeepWiFi can reliably identify channels for transmission and sustain its throughput. For 9 users operating over 40 channels, we show that there is no throughput loss for DeepWiFi when jamming likelihood is less than 60\% and there is only 14\% throughput loss when  jamming likelihood is 80\%. \reev{Finally, we show that DeepWiFi outperforms a jamming-resistant MAC protocol called Jamming Defense (JADE) \cite{Jade} and is robust against adaptive jammers.}
	
	\subsection{Related Work}
	\rev{Jamming and anti-jamming mechanisms with machine learning are studied in} \cite{Yi2018,Terpek18} \rev{in which an adversarial user builds a deep learning based classifier  to predict if there will be a successful transmission (replied with ACK messages) or not (no ACK) using an exploratory (inference) attack to jam data transmissions. A transmitter in return can launch a causative attack to poison jammer's classifier (by adding controlled perturbations to transmit decisions) as a defense mechanism. Similar to}  \cite{Yi2018,Terpek18}\rev{, DeepWiFi considers users with jamming capabilities. The difference is that DeepWiFi identifies jammed channels and enables transmissions (by adapting the MCS) even under jamming instead of backing off (which would be the case in the baseline WiFi). There is also a rich literature on modulation recognition studies that distinguish signals with different modulation schemes.} These related studies include earlier work in \cite{Gardner88,Azzouz} that use cyclic spectrum characteristics extracted from raw signal and more recent efforts in \cite{OShea2016,OShea2016b} that use the raw I/Q samples as inputs using CNN. \rev{ Since a frame may consist of multiple modulations (in the preamble and data portions) and modulations may change across short MAC frames (5.484~msec in IEEE 802.11ac \cite{Gast}), the problem in DeepWiFi is different from} \cite{OShea2016,OShea2016b} and focused on  classifying waveforms (chunked into frames) within the frame time rather than simple modulations that remain fixed over longer periods of time. \reev{Note that overall processing overhead for the front end processing, signal classification, and signal authentication of DeepWiFi is measured as $0.1546$~msec ($2.8\%$ of 802.11ac frame) to process each sample.} Furthermore, we note that the conventional energy detectors for spectrum sensing cannot be applied in our framework because they cannot distinguish busy and jamming signals such that an adversary can easily create a jamming signal with the same energy characteristics to fool an energy detector. 
	
	There are several characteristics that can be used for RF fingerprinting. One can focus on transient \cite{Ureten2007} or the steady-state \cite{Ramsey12,Candore09} behavior of the transmitted signal. For example, \cite{Ureten2007} studied the unique transient characteristics of a transmitter, which are often attributed to RF amplifier, frequency synthesizer modules, and modulator subsystems. The duration of the transient behavior changes depending on the type and model of the transmitter. Also, as a device ages, its transient behavior can change \cite{Ureten2007}. In DeepWiFi, we opt not to use transient behavior as a feature since it is highly susceptible to noise and interference effects and requires precise timing of when the signal transmission has started. Instead, DeepWiFi uses steady-state characteristics between a transmitter and a receiver pair such as frequency and timing synchronization offset, both of which are often attributed to the radio's oscillator, power amplifier, and digital-to-analog converter. There are two relevant studies that use steady-state characteristics similar to this paper. First, in \cite{Ramsey12}, three instantaneous signal characteristics such as the amplitude, phase, and frequency of the signal are used. After a preprocessing step (removal of the mean and normalization), their features (such as the variance, skewness, and kurtosis) are extracted within predefined window sizes \cite{Ramsey12}. Second, the frequency offset, magnitude and phase offset, distance vector, and I/Q origin offset are used in \cite{Candore09} as features for RF fingerprinting, and weighted voting classifiers and maximum likelihood classifier are developed for fingerprinting. In DeepWiFi, we use machine learning for signal authentication purposes. 
	
	Ad hoc networking over WiFi has two standards, namely, the ad hoc mode in 802.11 and its successor WiFi Direct. The ad hoc mode (IBSS) is one of the two modes of operation in the 802.11 standard, the other mode being the commonly used mode of infrastructure-based mode (IM-BSS). The IBSS mode enables direct communication between any devices without the need for an access point (AP). This standard has been evolved to what is known as WiFi Direct that is considered to be the main standard of ad hoc networking over WiFi, particularly, using Android smartphones \cite{WiFiAlliance}. Google has developed a peer-to-peer (P2P) framework, WiFi P2P, that complies with WiFi Direct standard. Devices after Android 4.0 (API level 14) can discover other devices and connect directly to each other via WiFi without an intermediate AP, when both devices support WiFi P2P \cite{AndroidDevelopers}. The APIs include calls for peer discovery, request, and connection, all of which are defined in the WifiP2PManager class (see \cite{AndroidDevelopers} for more details). A related work that uses WiFi direct in ad hoc mode and implements \rav{multi-hop} routing is \cite{Funai17}, which improves multi-group link connectivity with low overhead. \rav{Note that WiFi Direct, as defined in the standard, does not support multi-hop communications, but recent work such as \cite{Funai17} has shown that multi-hop routing can be implemented using WiFi Direct. With the four novel contributions of DeepWiFi focused on PHY and MAC layers (as highlighted in Section~\ref{sec:system}), DeepWiFi is designed to support both single-hop and multi-hop communications.} The routing approach in DeepWiFi differs from \cite{Funai17} such that routing based on the backpressure algorithm uses both queue and channel information. 
	
	\reev{The RF preprocessing step in DeepWiFi utilizes autoencoders \cite{DAE_Bengio_08,AutoencoderPascal2010}. In the literature, there are several studies that use autoencoders to extract features in a high dimensional space and provide a feature representation of the data, and then train a separate classifier using these features (e.g., \cite{KingmaSemiSupervised}). In this paper, we also use the same semi-supervised approach. Other prepossessing methods such as Short Time Fourier Transformation, Choi-Williams Transformation, and Gabor Wavelet Transformation have been used in \cite{Gunes2019} to preprocess I/Q data before running them through a CNN. In this paper, our preprocessing step is data-driven and consists of a denoising autoencoder that is shown to suppress noise in the input to the deep learning classifier and results in a small reconstruction loss.} 
	
	The rest of the paper is organized as follows. Section~\ref{sec:system} describes the system model. The DeepWiFi protocol overview is given in Section~\ref{sec:protocol}. RF front end processing of DeepWiFi is described in Section~ \ref{sec:step1}. This is followed by the description of spectrum sensing and signal classification of DeepWiFi in Section~\ref{sec:step2}. RF fingerprinting is applied in Section~\ref{sec:step3} to authenticate signals in DeepWiFi. Section~\ref{sec:step4} describes channel selection and channel access of DeepWiFi. It is followed by the descriptions of power control for LPI/LPD and MCS adaptation of DeepWiFi in Sections~\ref{sec:step5} and \ref{sec:step6}, respectively. Section~\ref{sec:step7} presents the routing extension of DeepWiFi. Simulation setting is described in Section~\ref{sec:simulation}. Performance results of DeepWiFi and baseline WiFi are presented in Section~\ref{sec:performance}. \rev{The implementation, overhead, and complexity aspects of DeepWiFi are discussed in Section~}\ref{sec:implementation}. Section~\ref{sec:conclusion} concludes the paper. 
	\section{System Model} \label{sec:system}
	\subsection{Network Setting}
	A wireless multi-hop network of $n$ users is considered. Each user may act as the source, the destination, or the relay for packet traffic. Each source $i$ generates unicast packets for traffic flow $s_{ij}$, addressed to destination $j$ with rate $r_{ij}$. \rev{While DeepWiFi algorithms do not require synchronization, we consider slotted time in discrete time simulation and follow the 802.11ac MAC's synchronization that uses the synchronizing timers in the preamble of a frame (Non-HT Training Field} \cite{Gast}). Each user $i$ holds a separate queue (with length $Q_i^s(t)$ at time slot $t$) for every packet flow $s$. Packets in each queue are served in a first-come-first-served (FCFS) fashion. Packets are allocated to 802.11ac MAC frames before transmission. If the queue length is smaller than the frame, zero padding is used for the missing data part. If the queue length is greater than the frame, packets in the queue are partitioned into multiple frames before transmission. There are $m$ channels available. At any given time slot, a user may select one of these channels to transmit packets. Users share these channels for data transmissions and control information exchanges. There is no centralized controller. \rev{Users make their own decisions in a distributed setting.} There are $n_J$ jammers that aim to interfere with transmissions. Without loss of generality, each jammer is assigned to one of the channels (i.e., $n_J = m$). \reev{There are three types of jammers:}
	\begin{itemize}
		\item \emph{Probabilistic (random) jammer}: The jammer is turned on with fixed jamming probability $p_J$ at any given time slot.
		\item \reev{\emph{Static sensing-based jammer}}: The jammer is turned on if it detects a signal on the channel using a constant sensing threshold $\tau$, else it is turned on with the probability of jamming $p_J$ at any given time slot.
		\item \reev{\emph{Adaptive jammer}: This jammer is a sensing-based jammer where channel sensing threshold is adaptively adjusted. If the jammer senses a signal on the channel, it is turned on.}
	\end{itemize}
	\reev{Both the static and adaptive jammers use channel sensing measurements to make jamming decisions. The sensing threshold is constant for the static sensing-based jammer, while the adaptive jammer changes the threshold by observing the actions of the transmitter in a given time window.}
	
	IEEE 802.11ac standard is followed for the PHY implementation and MAC frame format. The center frequency is 5.25~GHz. The instantaneous bandwidth is 20~MHz and each OFDM subcarrier occupies a bandwidth of 312.5~kHz. The subcarriers can be used independently and incoming data bits can be distributed among the subcarriers (possibly depending on the channel frequency response). Thus, a 20~MHz channel consists of 64 subcarriers. We allocate 8 subcarriers for pilot, 16~subcarriers as null, and 40~subcarriers for data (i.e., $m = 40$).
	
	Each user selects its MCS depending on its received SINR. \rev{The link rate from user $i$ to user $j$ at time slot $t$ is denoted by $c_{ij}(t)$ and depends on the selected MCS, transmit power, and consequently the observed SINR.} On each channel, the SINR at a user depends on the channel and interference effects among users (legitimate transmitters and jammers). Next, we describe how the channels among users and their signals are generated.
	\subsection{Channel and Waveform Data Generation} \label{sec:channel}
	We generate the physical signals and channels using the 802.11ac library in MATLAB WLAN System Toolbox. Several channel models are offered. Each channel model considers a different breakpoint distance, root mean square (RMS) delay spread, maximum delay, Rician K-factor for non-light of sight (NLOS) conditions, number of taps, and number of clusters.
	Waveform related parameters are specified to generate the signal such as the input data bits, number of packets, packet format (VHT, HT, etc.), idle time (added after each packet), scrambler initialization, the number of transmit antennas, payload length, MCS, and bandwidth using the \texttt{wlanWaveformGenerator} function of the MATLAB WLAN System Toolbox. This function generates a time-domain I/Q waveform $\bm{x}$ that goes through the channel, where small scale fading and other channel impairments are applied. The MATLAB function used for this operation is \texttt{wlanTGacChannel} and it performs an $\mathbf{H} \bm{x}$ operation, where $\mathbf{H}$ is the channel matrix. There are several parameters that can be set during the channel generation such as delay profile (models A-F are shown in Table \ref{table:delaymodel}), sample rate, carrier frequency, transmission direction (uplink or downlink), the number of receive and transmit antennas, spacing between antennas, and large-scale fading effects (`None', `Pathloss', `Shadowing', or `Pathloss and Shadowing'). Next, additive white Gaussian noise (AWGN) is added to the  signal that is passed through the channel using the \texttt{awgn} command of the MATLAB WLAN System Toolbox. Thus, we obtain the channel effect $\mathbf{H} \bm{x}+\bm{n}$ induced on the I/Q waveform $\bm{x}$.
	\begin{table}
		\caption{Delay models.}
		\centering
		{\small
			\begin{tabular}{c|c|c|c|c|c|c}
				\toprule
				& \multicolumn{6}{c}{\textbf{Model}} \\ 
				\textbf{Parameter} & A & B & C & D & E & F \\ \midrule 
				Breakpoint distance (m) & 5 & 5 & 5 & 10 & 20 & 30 \\ \hline
				RMS delay spread (ns) & 0 & 15 & 30 & 50 & 100 & 150 \\ \hline
				Maximum delay (ns) & 0 & 80 & 200 & 390 & 730 & 1050 \\ \hline
				Rician K-factor (dB) & 0 & 0 & 0 & 3 & 6 & 6 \\ \hline
				Number of taps & 1 & 9 & 14 & 18 & 18 & 18 \\ \hline
				Number of clusters & 1	& 2 & 2 & 3 & 4 & 6 \\
				\bottomrule
			\end{tabular}
		}
		\label{table:delaymodel}
	\end{table}
	
	\section{Overview of the Deep WiFi Protocol} \label{sec:protocol}
\begin{figure}
		\centering
		\includegraphics[width=\columnwidth]{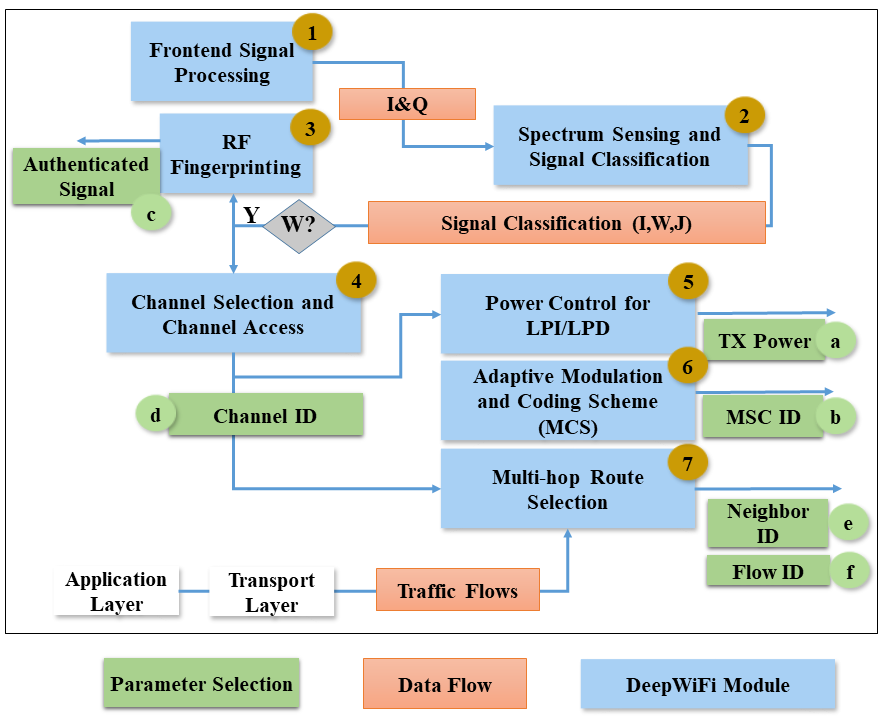}
		\caption{DeepWiFi protocol diagram with seven steps.}\label{fig:protocol3}
	\end{figure}

	On top of 802.11ac, each user individually and asynchronously runs the DeepWiFi protocol stack with the following seven main steps (shown in Fig.~\ref{fig:protocol3}):
	
	\begin{enumerate}
		\item \emph{RF front end processing}: Each user hops among WiFi channels (one by one), collects RF signal on each sensed channel and processes the RF signal at the RF front end to build the I/Q data and extract features.
		\item \emph{Spectrum sensing and signal classification}: Each user applies \emph{deep learning} to these features (the output of the RF front end) in order to classify (label) each channel as idle (I), jammed (J) or used by another WiFi device (W).
		\item \emph{Signal authentication}: Each user applies machine learning-based RF fingerprinting to authenticate legitimate WiFi signals at the physical layer.
		\item \emph{Channel selection and channel access}: Each user backs off on any busy channel used by legitimate WiFi signals (W) (to resolve future conflicts), selects an idle channel (I); if none, selects a jammed channel (J) (including channels used by the non-legitimated WiFi signals) with the best SINR for data transmission. \rev{The use of jammed channels (when no idle channel is available) corresponds to the degraded mode, where a non-zero throughput can be still achieved.}	
		\item \emph{Power control for LPI/LPD}: \rev{Each user selects the transmit power below the jammer threshold level to avoid detection by jammers and achieve LPI/LPD.}
		\item \emph{Adaptive modulation and coding}: \rev{There are nine possible MCS options to choose from in 802.11ac. Each user selects the best MCS based on the measured SINR to maximize the achievable rate on the selected channel.}
		\item \emph{Routing}: \rev{Each user makes the routing decision by selecting the flow to serve and the next hop for transmission by applying the \emph{backpressure algorithm}, which optimizes a spectrum utility depending on traffic congestion and link rate (computed in Step 6). Lower layers of DeepWiFi are transparent to the routing algorithm and can be combined with other efforts for multi-hop networking such as the extension of WiFi Direct for multi-group networking} \cite{Funai}.
	\end{enumerate}

	The output of the DeepWiFi protocol is specified as:
	\begin{enumerate}[\:\: a.]
		\item \emph{TX power} specifies the transmit power.
		\item \emph{MCS ID} specifies which  MCS of IEEE 802.11ac is used and its corresponding rate.
		\item \emph{Authenticated signal} specifies which signals belong to legitimate WiFi. 
		\item \emph{Channel ID} specifies  which channel to send the next data packet.
		\item \emph{Neighbor ID} specifies which neighbor to select for the next hop transmission.
		\item \emph{Flow ID} specifies  which application traffic flow to serve when the data packet is transmitted.
	\end{enumerate}
	
	These outputs tune parameters of the 802.11ac network protocol stack at each user. Specifically, outputs $a$-$c$ tune the physical layer, outputs $d$ and $e$ tune the link/MAC layer, and outputs $f$ and $g$ tune the network layer, as shown in Fig.~\ref{fig:protocol3}. 
	\reev{\section{DeepWiFi Protocol Steps}}
	\subsection{Step 1: RF Front End Processing} \label{sec:step1}
	\emph{Input}: The RF signal.
	\\
	\emph{Output}: The set of extracted features given to Step 2 and the I/Q data given to Step 3.
	
\begin{figure}
		\centering
		\includegraphics[width=\columnwidth]{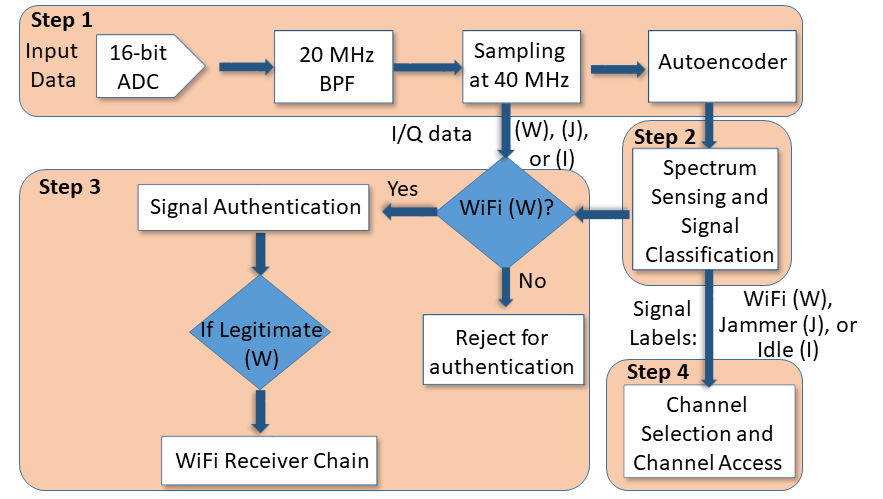}
		\caption{RF front end processing of DeepWiFi (Step 1) and its connections to Steps 2 and 3.}\label{fig:frontend}
	\end{figure}

	As shown in Fig.~\ref{fig:frontend}, the following steps are pursued to process the RF signal.
	\begin{enumerate}
		\item 16-bit analog-to-digital converter (ADC) is used to sample the signal by allocating 8 bits for real and imaginary parts of the signal.
		\item The digitized signal is bandpass filtered with 20MHz instantaneous bandwidth to remove interference from adjacent bands.
		\item The digitized and bandpass-filtered signal is sampled at 40MHz.
		\item The deep-learning based autoencoder takes the input samples and reduces them into latent features. 
	\end{enumerate}
	
	RF front end processing provides the I/Q data to Step 3 (signal authentication) and the reduced set of features to Step 2 (spectrum sensing and signal classification). The details of this procedure (shown in Fig.~\ref{fig:protocol3}) are given below:
	\begin{enumerate}
		\item Spectrum sensing and signal classification (Step 2) takes the  outputs of the  autoencoder in Step 1 and classifies the input signal as WiFi (W), Jammer (J), or Idle (I). 
		\item A logic takes the signal classification result as input. If the signal is a WiFi signal, it passes the I/Q data to the physical layer authentication (RF fingerprinting) module, which constitutes the first step of the physical layer security. 
	\end{enumerate}
	
	\reev{Instead of using I/Q data directly as input to the signal classifier, we use a denoising autoencoder to extract the features of the received signal that are then fed into the signal classifier. In an RF environment, we typically have many unlabeled data samples (that can be collected by sensing the spectrum) while the number of labeled data samples (needed to train the classifier) is relatively small compared to the dimension of each data sample (40K is the dimension per sample in our case). However, unlabeled data can be used to train an autoencoder that reduces the dimension of the input data to the classifier to build a reliable classifier from few labeled data samples. In addition, we use a denoising autoencoder that further suppresses the noise in the input. Another benefit of the autoencoder may be observed when the environment changes (e.g., new channel conditions (distributions) or unknown signals passing through known or unknown channel distributions). An autoencoder trained with unlabeled data in the new environment may help adapt to the effects of this new environment by robust feature extraction. Then, a small number of labeled data samples in the new environment could be sufficient. These capabilities would not be achievable if we were to  use the I/Q data directly as input to the deep neural network-based classifier as we cannot train a reliable classifier for a high-dimensional problem using a small number of labeled training samples.}
	
	To preprocess the data, we also tried different standard techniques such as Principal Component Analysis (PCA) \cite{PCA} and t-distributed stochastic neighbor embedding (tSNE) \cite{tSNE}. However, they failed in separating signals of interest. These results are reported in the Appendix.  Instead, DeepWiFi uses an autoencoder to extract features from the I/Q data. An autoencoder is a deep neural network that is trained to reconstruct its input and consists of two neural networks, namely, an \emph{encoder} $\bm{h} = f_{\theta} (\bm{x})$ and a \emph{decoder} that produces a reconstruction $\bm{r} = g_{\Phi}(\bm{h})$, where $\theta$ is the set of weights and biases of the neural network corresponding to the encoder and $\Phi$ represents that of the decoder. The neural networks $f$ and $g$ can be constructed as FNN or CNN. Autoencoder is used for unsupervised learning of efficient codings. DeepWiFi uses the autoencoder to learn a representation (encoding) for a set of data, for the purpose of feature learning and dimensionality reduction. In particular, DeepWiFi uses a \emph{denoising autoencoder} that adds noise to its inputs and trains it to recover the original noise-free inputs. This method prevents the autoencoder from trivially copying its inputs to its outputs and the autoencoder finds the patterns in the data, while avoiding overfitting. 
	
	The preprocessing of DeepWiFi (ADC, bandpass filtering, and sampling) produces the I/Q data that has the dimension of 40000 (20000 for I and 20000 for Q components) for each time instant in Step 1. DeepWiFi applies denoising autoencoder to this I/Q data, and determines the latent features that are further fed to the signal classifier of DeepWiFi.

	The denoising autoencoder of DeepWiFi adds the Gaussian noise to the I/Q data (to prevent overfitting) and then applies four hidden layers (the first two layers for encoding and the last two layers for decoding) after one initial normalization layer. Hidden layers are trained through the \emph{backpropagation algorithm} to minimize the minimum squared error (MSE) as the loss function. We used the hyperbolic tangent function (tanh) as the activation function that performs $f(x)=\mathrm{tanh}(x)$ operation. In a neural network, the activation function is used as an abstraction representing the rate of action potential firing in the cell.  We performed hyperparameter optimization and observed that a Gaussian noise $N$ with zero mean and variance of 0.1 gives the best reconstruction loss and avoids overfitting. DeepWiFi uses the following autoencoder structure: 
	\begin{itemize}
		\item Hidden layer 1: FNN with 534 neurons and \reev{tanh} activation.
		\item Hidden layer 2: FNN with 66 neurons and \reev{tanh} activation.
		\item Hidden layer 3: FNN with 534 neurons and \reev{tanh} activation.
		\item Hidden layer 4: FNN with 40000 neurons and \reev{tanh} activation.
	\end{itemize}
	The input and output layers have the same dimension.

\begin{figure}
		\centering
		\includegraphics[width=0.8\columnwidth]{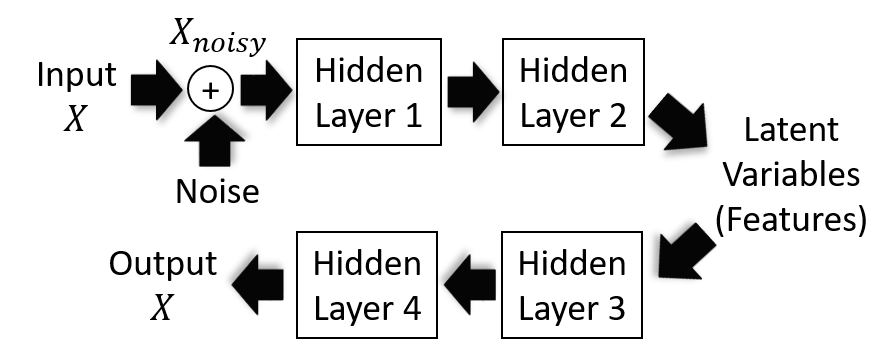}
		\caption{Denoising autoencoder in DeepWiFi.}\label{fig:dae}
	\end{figure}

	A denoising autoencoder adds a Gaussian noise $N$ with $0$ mean and variance of $\sigma^2$ to the input data samples $X$. The resulting data $X_{noisy} = X + N$ is input to the neural network. Denoising autoencoder solves the following loss function:
	\begin{align}
	(\Phi^\ast,\theta^\ast) = \min_{(\Phi,\theta)} E_X [||g_{\Phi}(f_{\theta}(X_{noisy})) - X||^2 ],
	\end{align}
	where $E_X[\cdot]$ denotes the expectation over $X$ and \rev{$||\cdot||^2$ is the $\ell_2$-norm (the Euclidean norm)}. The structure of denoising autoencoder is shown in Fig.~\ref{fig:dae}.
	
\begin{figure}[t!]
		\centering
		\includegraphics[width=0.9\columnwidth]{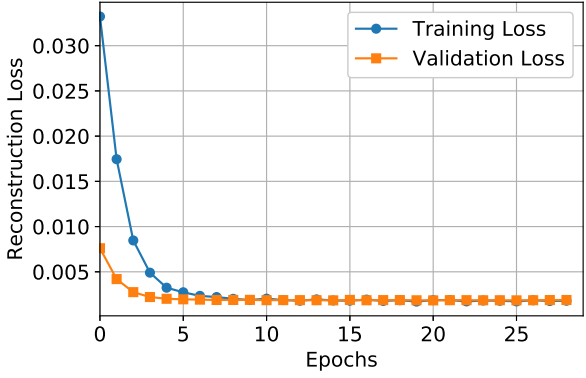}
		\caption{Training and test losses during denoising autoencoder training.}\label{fig:dae_loss}
	\end{figure}

	We implemented a denoising autoencoder using the TensorFlow framework, which takes the input I/Q data and reduces its dimensions. These features represent the latent variables (extracted features). The reconstruction loss between the original signal $X$ and the reconstructed signal $\hat{X}$ is computed by $E[||X-\hat{X}||^2]$. To find the best parameters, we performed a hyperparameter optimization using the Hyperband framework \cite{Hyperband}. This framework can provide over an order-of-magnitude speedup over Bayesian hyperparameter optimization methods. The Hyperband randomly samples a set of hyperparameter configurations, evaluates the performances of all current configurations, sorts the score of configurations, and removes the worst scoring configurations (successive halving). The process is repeated progressively with increasing number of iterations. Therefore, only the configurations that yield good results are evaluated thoroughly. Note that we use Hyperband to optimize the parameters of neural networks described in Sections~\ref{sec:step1}-\ref{sec:step2}. We considered the maximum amount of resources that can be allocated to a single configuration as $81$ iterations and a downsampling rate of three ($\eta=3$).
	The dataset used in the training of the denoising autoencoder consists of equal number of WiFi, jammer, and noise signals. The dataset is divided into 80\% and 20\% splits for the training and test sets, respectively. We use a batch size of 64 samples. Fig.~\ref{fig:dae_loss} depicts the loss in the training and test sets during the training process. We observe that the loss gradually decreases in both sets and no overfitting is observed.
	
\begin{figure}[t!]
		\centering
		\includegraphics[width=\columnwidth]{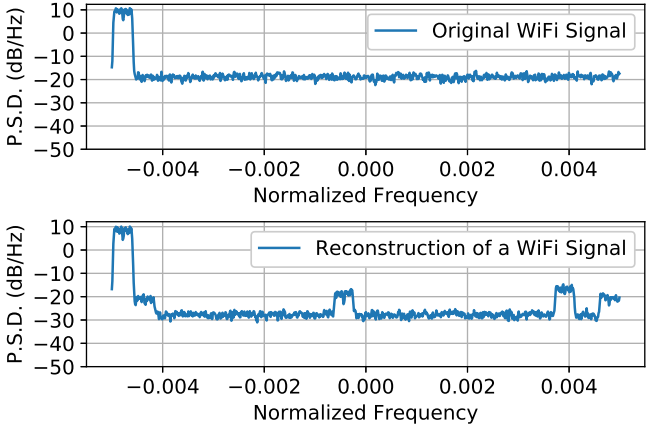}
		\caption{Reconstruction of a WiFi signal in the test set.}\label{fig:dae_wifi_reconstuction}
	\end{figure}

	Fig.~\ref{fig:dae_wifi_reconstuction} illustrates the legitimate WiFi signal in the test set and its reconstruction. We observe the noise suppression on the sidebands up to 10~dB. This highlights the effect of denoising autoencoder on the noise reduction. Note that a similar denoising effect using denoising autoencoders has been reported in the image processing \cite{DAEXie} and Fig.~\ref{fig:dae_wifi_reconstuction} shows the same effect in the RF domain.

	The output of the autoencoder is the set of extracted features that are given to the classifier in Step 2.
	
	\subsection{Step 2: Spectrum Sensing and Signal Classification} \label{sec:step2}
	\emph{Input}: The I/Q data on the sensed channel from Step 1.
	\\
	\emph{Output}: The classification of the channel to idle, used by WiFi or jammed, given as input to Step 3. 
	
	Each user applies a deep neural network-based classifier to the received I/Q data on the sensed channel and classifies the captured signal as noise (idle), WiFi signal, or jammer signal. Features are the I/Q data received over time (output of Step 1). There are three potential labels assigned to each channel: idle (I), another WiFi device (W), and jammed (J). Training data is collected offline and training is performed offline. The trained classifier is loaded to each user offline. Only neural network weights and biases are stored in the memory. Each radio individually runs its classifier online.
	
	\subsubsection{Signal Classes (I), (W) and (J)}
	
	For training and testing phases, signals of each label are generated as follows:
	
	\begin{enumerate}
		\item \emph{Noise}: The background noise for WiFi is between $-80$~dB and $-100$~dB. To emulate such a case, we choose a number uniformly at random between $-80$~dB and $-100$~dB in the frequency domain and then take the inverse Fourier transform to obtain the time samples. The output is stored in the training data as the background noise samples with label (I).
		
		\item \emph{WiFi Legitimate Signal}: We use the MATLAB WLAN System Toolbox to generate the WiFi signal including the preamble and payload. We consider the following parameters:
		\begin{enumerate}
			\item Center frequency: 5.25 GHz,
			\item Channel Bandwidth: 20 MHz,
			\item TX-RX distance: 5m (default value; we vary the SINR in simulations),
			\item Normalize Path Gains: True,
			\item Transmit antenna spacing: 0.5 wavelength (default),
			\item Receive antenna spacing: 0.5 wavelength (default),
			\item Packet format: VHT,
			\item Scrambler initialization: 93 (default),
			\item Channel coding: BCC (binary convolutional coding),
			\item APEP Length: 1,
			\item PSDU Length: 36 (default; PSDU length is the number of bytes carried in the user payload. For a single user, the PSDU length is scalar integer from $1$ to $2^{20} -1$),
			\item Long guard interval length: 800 ns,
			\item Short guard interval length: 400 ns,
			\item A-MPDU Length: 256 bytes,
			\item Number of transmit antennas: 1,
			\item Number of spatial streams transmitted: 1, and
			\item MCS: Varying between 0-9.
		\end{enumerate}
		
		We generate channel coefficients using the specific channel model (as discussed in Section~\ref{sec:channel}) and pass the WiFi signal through the channel. Then, we add white Gaussian noise. The output is stored in the training data as the WiFi signal samples with label (W).
		
		\item \emph{Jammer Signal}: In the time domain, we first generate normally distributed random numbers with zero mean and variance of one. Then we up-sample these samples to a selected carrier with a bandwidth that is the same as the WiFi signal. We generate channel coefficients using the same channel models discussed in Section~\ref{sec:channel} and pass the jammer signal through the channel. Then, we add white Gaussian noise. The output is stored in the training data as the jammer signal samples with label (J).
	\end{enumerate}
	
	12000 samples are generated using different channel models. 80\% of data is used for training and 20\% is allocated to testing. Training data set includes signals generated under six different channel conditions (2000 samples per channel model).
	
	\subsubsection{Simple Machine Learning Classifiers}
	For benchmark comparison, we evaluate the performance of SVM classifier in signal classification. As its performance depends on the kernel type and hyperparameters used, the hyperparameters are tuned for the best accuracy. Two of the most common kernels, linear kernel and radial basis function (RBF) kernel, are used. For the RBF kernel, there are two parameters that are  considered, $C$ and $\gamma$. The parameter $C$ is common in all SVM kernel types and it trades off misclassification of training examples against simplicity of decision surface. When $C$ is a small number, the decision surface is smoother, and a high $C$ puts more emphasis on classifying all training examples correctly. So $C$ trades error penalty for stability. The second parameter $\gamma$ is the free parameter of Gaussian RBF, $K(x_i,x_j) = \exp(-\gamma||x_i - x_j||^2)) $ where $\gamma>0$. It defines the influence of each training example. A larger $\gamma$ has smaller variance affecting only the closer neighbors. For hyperparameter tuning, $C = [0.0001, 0.001, 0.01, 0.1, 1.0, 10.0, 100.0, 1000.0]$ is used for linear kernel, and $\gamma$ with the same $C$ values for the Gaussian RBF kernel. In the evaluation step, we implement $k$-fold cross-validation to calculate cross-validation accuracy and set $k$ as 10. Using a grid search, we obtain the score of best performing model. The best performance is achieved using RBF kernel with $C = 1.0$ and $\gamma=0.1$, which achieves only 66\% accuracy.

	\subsubsection{Deep Learning Classifiers}
	We design two deep neural networks architectures, FNN and CNN, for the signal classification task that achieves small memory footprint, high accuracy, and low inference time. These deep neural networks are implemented in TensorFlow  using the \texttt{Keras} library \cite{Tensorflow}. Backpropagation is applied to train the neural network using a cross entropy loss function that is defined as
	$\mathcal{L} = -\sum_{i=1}^m \beta_i \log(y_i)$, where $\boldsymbol{\beta} = \{ \beta_i \}_{i=1}^m$ is a binary indicator of ground truth such that for a sample from label $k$, $\beta_k=1$ whereas the other entries are all zeros. The neural network prediction is denoted by $\textbf{y} = \{y_i\}_{i=1}^m$. In both architectures, we used the ADAM optimizer \cite{Adam} with the learning rate of $10^{-5}$.

	\subsubsection{FNN}\label{Section:FNN}
	We tune the hyperparameters of the FNN (such as the number of layers, number of neurons per layer and activation functions). We use rectified linear unit (ReLU) as the \emph{activation function} on neural network layer outputs that performs the $f(x) = \max(0,x)$ operation. The advantages of using ReLU activation function is that it is a bit faster to compute than other activation functions in hardware and its does not suffer from vanishing gradient problem since it does not saturate for positive values, as opposed to the logistic function and hyperbolic tangent function \cite{Geron}. The resulting FNN consists of the following layers:
	\begin{itemize}
		\item Fully connected layer with 15 neurons (ReLU activation).
		\item Dropout layer with 50\% dropout probability.
		\item Fully connected layer with 3 neurons (Softmax activation).
	\end{itemize}
	It uses the \emph{dropout layer} to drop out some neuron outputs from the previous layer of a neural network and serves the purpose of regularization for reducing overfitting by preventing complex co-adaptations on training data. For the output layer, \emph{softmax activation} function $f_i(\bm{x}) = e^{x_i}/(\sum_{j}e^{x_j})$ is  used at the final layer of the neural network. Softmax activation function is generally a good choice for classification tasks for which output classes are mutually exclusive. FNN achieves the average accuracy of $98.75\%$ in predicting the correct signal labels. The confusion matrix for FNN is shown in Table \ref{table:fnn_cm}. 
	
	\begin{table}
		\caption{Confusion matrix for the FNN.}
		{\small
			\begin{tabular}{rc|c|c|c|}
				\multicolumn{2}{r}{} & \multicolumn{3}{c}{\textbf{Predicted Label}} \\ \\
				\multicolumn{2}{r}{} &  \multicolumn{1}{c}{(I)} &  \multicolumn{1}{c}{(W)} & \multicolumn{1}{c}{(J)} \\  \cline{3-5}
				& (I) & 35.6\% & 0.0\% & 0.0\% \\ \cline{3-5}
				\textbf{True Label} & (W) & 0.0\% & 32.4\% & 0.5\% \\ \cline{3-5}
				& (J) & 0.1\% & 0.75\% & 30.75\% \\ \cline{3-5}
		\end{tabular}}
		\label{table:fnn_cm}
	\end{table}

	\subsubsection{CNN} We tune the hyperparameters of the CNN. The resulting CNN consists of the following layers:
	\begin{itemize}
		\item Five cascades of the following layers concatenated:
		\begin{itemize}
			\item A 2-D convolutional layer with 32 filters and kernel sizes of (2,5).
			\item A 2-D convolutional layer with 32 filters and kernel sizes of (2,5).
			\item Max pooling layer with kernel size of (2,2) and stride of (2,2).
			\item Batch normalization layer.
		\end{itemize}
		\item Fully connected layer with 18 neurons.
		\item Dropout layer with a $50\%$ dropout probability.		
		\item Fully connected layer with 3 neurons.
	\end{itemize}
	
	The \emph{2-D convolutional layer} is used to apply sliding filters to the input. This layer convolves the input by moving the filters along the input vertically and horizontally, computing the dot product of the weights and the input, and then adding a bias term. The \emph{max pooling layer} is used to progressively reduce the spatial size of the representation to reduce the number of parameters and amount of computation in the network, and hence to control overfitting. The max pooling layer performs down-sampling by dividing the input into rectangular pooling regions, and computing the maximum of each region. The \emph{batch normalization layer} is used to normalize each input channel across a mini-batch. The purpose is to speed up the training of CNN and reduce the sensitivity to network initialization. CNN achieves the average accuracy of $98.0\%$ in predicting the signal labels. The confusion matrix for CNN is shown in Table~\ref{table:cnn_cm}. 
	
	\begin{table}
		\caption{Confusion matrix for the CNN.}
		{\small
			\begin{tabular}{rc|c|c|c|}
				\multicolumn{2}{r}{} & \multicolumn{3}{c}{\textbf{Predicted Label}} \\ \\
				\multicolumn{2}{r}{} &  \multicolumn{1}{c}{(I)} &  \multicolumn{1}{c}{(W)} & \multicolumn{1}{c}{(J)} \\  \cline{3-5}
				& (I) & 35.6\% & 0\% & 0\% \\ \cline{3-5}
				\textbf{True Label} & (W) & 0.1 \% & 31.6\% & 1.0\% \\ \cline{3-5}
				& (J) & 0\% & 0.9 \% & 30.8\% \\ \cline{3-5}
			\end{tabular}
		}
		\label{table:cnn_cm}
	\end{table}
	
	\subsection{Step 3: Signal Authentication} \label{sec:step3}
	\emph{Input}: I/Q data from RF front end processing (Step 1) and channel label from spectrum sensing and signal classification (Step 2).
	\\
	\emph{Output}: Set of authorized signals (to be given to the WiFi receiver chain).
	
	The goal of physical layer authentication (illustrated in Fig.~\ref{fig:frontend}) is to augment the standard WiFi security (at Layer 2) at the physical layer by providing \emph{physical layer fingerprinting} capability for processing I/Q data and authenticate legitimate users. RF fingerprinting is motivated by mitigating replay attacks in wireless networks. Recent open-source firmware patches \cite{nexmon} enable the capability to store and transmit I/Q data in buffer. A software-defined radio (SDR) can also perform replay attacks by listening to the legitimate communication between two parties and replaying the I/Q data with the legitimate WiFi characteristics. In such a case, signal classification may not be enough for signal authentication, and RF characteristics in the signal transmitted by an adversary can be used to authenticate the signal in the PHY layer. Note that this step uses the I/Q data directly, not the features extracted by the autoencoder, as hardware impairments used in the authentication may be lost at the autoencoder. Therefore, I/Q data is directly used in this step. 
	The details of this authentication procedure (shown in Fig.~\ref{fig:frontend}) are given below:
	\begin{enumerate}
		\item Spectrum sensing and signal classification (Step 2) takes the outputs of the autoencoder in Step 1 and classifies the input signal as WiFi (W), Jammer (J), or Idle (I). 
		\item A logic takes the signal classification result as input. If the signal is a WiFi signal, it passes the I/Q data to the physical layer authentication (RF fingerprinting), which constitutes the first step of the physical layer security. 
		\item The physical layer authentication uses the physical layer impairments that are inherent in each transmitter and authorizes if it detects that the received signal comes from a legitimate transmitter. 
		\item If the received signal is authorized, the signal is processed by the WiFi receiver chain. 
	\end{enumerate}
	
	The objective is to analyze RF signal characteristics to authenticate legitimate WiFi users. One way of RF fingerprinting is based on location identification by capturing channel specific features (such as the Received Signal Strength Indicator (RSSI) levels). However, this approach does not effectively apply to mobile users with rapidly changing RSSI levels. Another way of RF fingerprinting is based on identification using radio characteristics that are divided into waveform domain and hardware domain characteristics (impairments). The waveform domain approach identifies transient-based behavior that lasts for a very short period of time (microseconds) and is hard to model. The transient characteristics are also prone to noise and interference effects. The hardware domain approach is based on capturing hardware impairments such as frequency, magnitude and phase errors, I/Q offset and synchronization offsets in the time and frequency domains. We pursue this second approach.
	
\begin{figure}
		\centering
		\includegraphics[width=1\columnwidth]{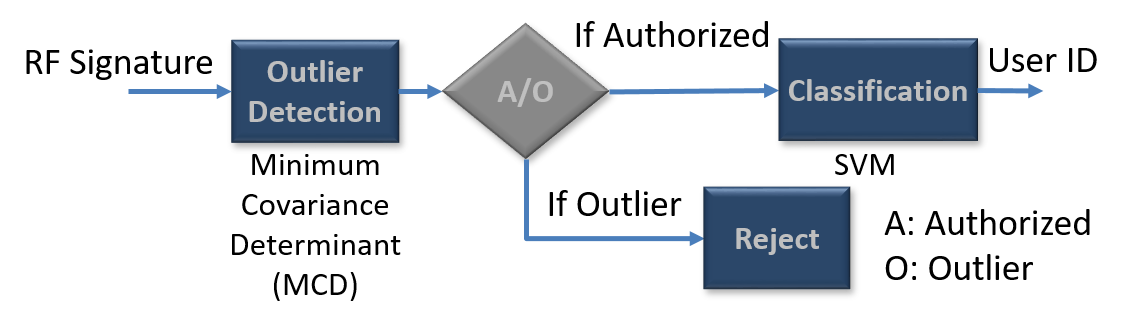}
		\caption{Two-layer approach for RF fingerprinting.}\label{fig:rf_fingerprint_two_step}
	\end{figure}

	To authenticate a signature, we use a two-layer approach (as shown in Fig.~ \ref{fig:rf_fingerprint_two_step}):
	\begin{enumerate}
		\item \emph{Outlier detection} determines if the signature belongs to any signature that is authenticated or not. If the signature is not authenticated, it is rejected. If the signature belongs to an authenticated user, then we proceed to next step.  
		\item \emph{Classification} validates the signature belongs to the transmitter that it claims. The classifier returns a user ID that is stored. When the data preamble is decoded, we verify that the sender and the output of the classifier (user ID) match. 
	\end{enumerate}
	In what follows, we describe the PHY-layer impairment model, feature extraction, data generation, and authentication steps. 
	
	\subsubsection{Modeling the PHY-layer impairments}
\begin{figure}
		\centering
		\includegraphics[width=0.8\columnwidth]{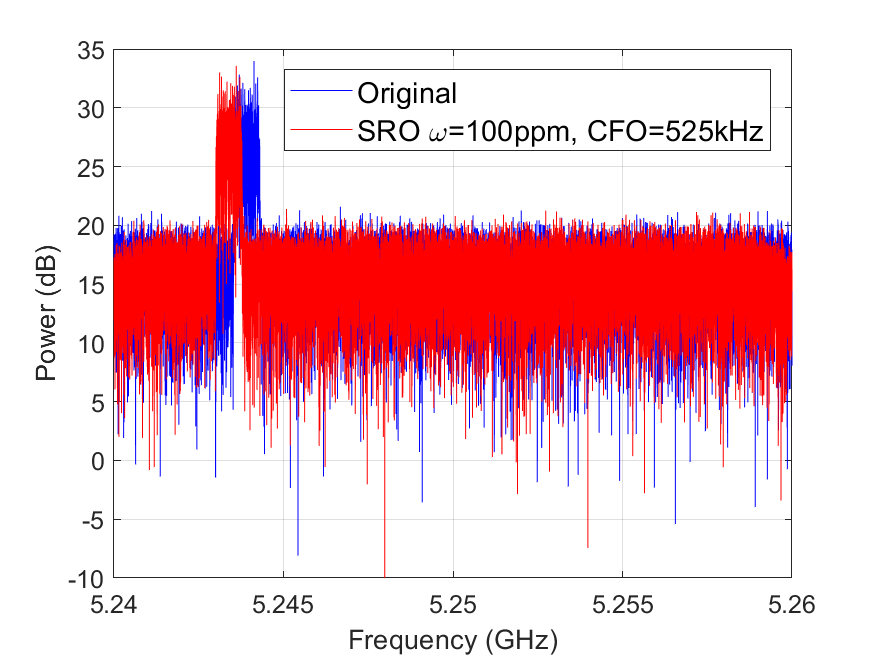} 
		\caption{WiFi signal with and without hardware impairments.}\label{fig:signal_sample}
	\end{figure}

	Training data is generated in MATLAB to account for different hardware impairments. In particular, the sampling rate offset, carrier frequency offset, and I/Q balance offset (amplitude and phase) are taken into consideration. To model the sample rate offset (SRO) between the transmitter and receiver, the transmitted waveform is resampled with a factor of $p/q$ times the original sample rate where $p$ and $q$ are the interpolation and decimation factors. These parameters are varied for each transmitter-receiver pair. We use the \texttt{resample} function of MATLAB. A frequency offset is introduced to the previous signal using the \texttt{helperFrequencyOffset} function of MATLAB. This joint SRO and carrier frequency offset impairments are suggested in \cite{matlab:JointSampleRate}. Finally, for each transmitter-receiver pair, a constant I/Q offset is added by an amount of $\psi$ for amplitude imbalance and $\phi$ for phase imbalance using the following formulation \cite{IQoffset}:
	\begin{align}
	&\kappa_I = 10^{0.5 \psi/20} \\
	&\kappa_Q = 10^{-0.5 \psi/20} \\
	&imb_I = {\rm I\!Re}\{txSig) \cdot  \kappa_I \cdot \exp(-0.5i \cdot \phi \pi/180) \\
	&imb_Q = {\rm I\!Im}(txSig) \cdot \kappa_Q \cdot \exp(i(\pi/2 + 0.5\phi \pi/180)) \\
	&rxSig = imb_I + imb_Q,
	\end{align}
	where $txSig$ and $rxSig$ are the transmitted and received signals, respectively, $\kappa_I$ and $\kappa_Q$ are the in-phase and quadrature gains, respectively, $imb_I$ and $imb_Q$ are the in-phase and quadrature imbalances, respectively, the amplitude imbalance $\psi$ is given in dB, and phase imbalance $\phi$ is given in degrees. An example is shown in Fig.~\ref{fig:signal_sample}, where SRO is denoted by $\omega$ and $\mathrm{CRO}$ is the carrier frequency offset.
	
	\subsubsection{Extracting the features from signal waveform}
	Next, we extract the signal features and obtain a fingerprint signature of each waveform. We first detect the I/Q imbalance using the MATLAB Communication Toolbox's \texttt{comm.IQImbalanceCompensator} and obtain a compensator coefficient \texttt{IQComp}. The I/Q coefficients are estimated using \texttt{iqcoef2imbal}, which computes the amplitude and phase imbalances given a compensator coefficient. 
	We also need to extract the synchronization in frequency and time domains. The frequency-domain synchronization has two components called the coarse and fine offsets. The time domain synchronization has only one component. We use \texttt{wlanCoarseCFOEstimate}, \texttt{wlanFineCFOEstimate}, and \texttt{wlanSymbolTimingEstimate} functions of MATLAB to estimate the coarse and fine center frequency offset, and symbol timing synchronization, respectively. Through these steps, we obtain five signatures: Coarse center frequency offset, fine center frequency offset, symbol timing synchronization, amplitude imbalance, and phase imbalance. Note that to extract these features, we have used domain knowledge rather than following a purely data-driven approach. This is due to the fact that hardware impairments introduce very subtle changes in the received signal such that domain knowledge is required to provide a robust algorithm.
	
	\subsubsection{Data generation}
	As a comprehensive dataset, we generate $50$ signal samples at SNR values from $5$ to $25$~dB in $5$~dB increments for each subband. The same experiment is repeated for $10$ users. To introduce the frequency offset impairments, we consider the SRO between a transmitter and receiver pair $j$ to be a multiple of 100 parts per million (PPM), i.e. $\omega_j = 100 j$ PPM as suggested in \cite{matlab:JointSampleRate}. For this purpose, we take the interpolation factor as $p = 10^{4}$ and vary the decimation factor $q=p - j$ for each user $j$ corresponding to a SRO of user $j$ as $\omega_j = (1 - p/(p-j))10^6$ PPM. Note that if the offsets are larger, RF fingerprint signatures will be easier to differentiate. These offsets are kept constant throughout different SNR and channel realizations. To introduce amplitude and phase imbalance,  $1$~dB and $10$ degrees increments are added between users such that $j$th user has $j$~dB and $10\times j$ degrees imbalance.

	\subsubsection{Outlier detection}
	To authorize the signals, we first use a simple outlier detection method in which the objective is to identify which signatures are authorized and which are not. We test and evaluate the performance of three different methods: (i) \emph{one-class SVM} \cite{oneclassSVM}, (ii) \emph{isolation forest} \cite{isolationForest}, and (iii) \emph{Minimum Covariance Determinant} (MCD) \cite{MCD,MCD2} methods. Our results show that MCD, also known as the elliptic envelope method, outperforms the other two methods. In what follows, we briefly describe all three methods and then present the performance evaluation results. 
	\begin{itemize} \item \emph{One-class SVM} is an \emph{unsupervised outlier detection} method. It learns a frontier delimiting the contour of initial observations. If a new observation lays within the frontier-delimited subspace, it is as coming from the same population (an authorized signature). Otherwise, if they lay outside the frontier, it is an outlier (an unauthorized signature). The parameter $\nu$ is the margin of the One-Class SVM that determines the probability of finding a new, but regular, observation outside the frontier.
		\item \emph{Isolation forest} isolates observations by randomly selecting a feature and then randomly selecting a split value between the maximum and minimum values of the selected feature. This recursive partitioning can be represented by a tree structure and the number of splittings required to isolate a sample is equivalent to the path length from the root user to the terminating user. This path length, averaged over a forest of such random trees, is a measure of normality and our decision function.
		\item \emph{MCD} fits an elliptic envelope such that any data point outside the ellipse is an outlier. It is a highly robust estimator of multivariate distributions \cite{MCD}. It uses the Mahalanobis distance such that $MD(\bm{x})= \sqrt{(\bm{x}-\boldsymbol{\mu}_x)^T \bm{\Sigma}^{-1}_x (\bm{x}-\boldsymbol{\mu}_x)}$, where $\boldsymbol{\mu}_x$ and $\bm{\Sigma}^{-1}_x$ are the mean and covariance of data $\bm{x}$. 
	\end{itemize}
	
	Suppose there are ten signatures, six of which are authorized and four are unauthorized (unseen) signatures. Table~\ref{table:cm_mcd} presents the confusion matrix of MCD outlier detection that achieves the average accuracy of 89.8\%. We also present the confusion matrix of one-class SVM with $\nu = 0.2$ and isolation forest methods in Tables~\ref{table:cm_onesvm} and \ref{table:cm_isolationForest}, respectively. The two classifiers achieve only 70.0\% and 69.1\% accuracy, respectively, in the same order as before. In Step 2, we use supervised classification and employ RBF-kernel based SVM with $C=1$ and $\gamma=0.1$ parameters, which achieves 100\% accuracy.
	
	\begin{table}[t!]
		\caption{Confusion matrix for MCD. ``A" stands for authenticated and ``O" stands for outlier.} 	\label{table:cm_mcd}
		{\small
			\begin{tabular}{rc|c|c|}
				\multicolumn{2}{r}{} & \multicolumn{2}{c}{\textbf{Predicted Label}} \\ \\
				\multicolumn{2}{r}{} &  \multicolumn{1}{c}{(A)} &  \multicolumn{1}{c}{(O)} \\  \cline{3-4}
				& (A) & 29.3\% & 10.2\% \\ \cline{3-4}
				\textbf{True Label} & (O) & 0\% & 60.5\% \\ \cline{3-4}
		\end{tabular}}
	\end{table}

	\begin{table}[t!]
		\caption{Confusion matrix for the one-class SVM method.} 	\label{table:cm_onesvm}
		{\small
			\begin{tabular}{rc|c|c|}
				\multicolumn{2}{r}{} & \multicolumn{2}{c}{\textbf{Predicted Label}} \\ \\
				\multicolumn{2}{r}{} &  \multicolumn{1}{c}{(A)} &  \multicolumn{1}{c}{(O)} \\  \cline{3-4}
				& (A) & 22.8\% & 16.7\% \\ \cline{3-4}
				\textbf{True Label} & (O) & 13.3\% & 47.2\% \\ \cline{3-4}
		\end{tabular}}
	\end{table}

	\begin{table}[t!]
		\caption{Confusion matrix for the Isolation Forest outlier detection method.} 	\label{table:cm_isolationForest}
		{\small
			\begin{tabular}{rc|c|c|}
				\multicolumn{2}{r}{} & \multicolumn{2}{c}{\textbf{Predicted Label}} \\ \\
				\multicolumn{2}{r}{} &  \multicolumn{1}{c}{(A)} &  \multicolumn{1}{c}{(O)} \\  \cline{3-4}
				& (A) & 15.1\% & 24.4\% \\ \cline{3-4}
				\textbf{True Label} & (O) & 6.5\% & 54.0\% \\ \cline{3-4}
		\end{tabular}}
	\end{table}

	\subsection{Step 4: Channel Selection and Channel Access} \label{sec:step4}
	\emph{Input}: The set of channel labels and SINR from Step 2.
	\\
	\emph{Output}: The ID of the channel selected for transmission.
	
	Each user individually starts scanning the set of $m$ channels (with a random initialization) and classifies each channel as (I), (W) or (J) with the deep learning-based classifier (as described in Step 3).
	\begin{enumerate}
		\item If channel $i$ is classified as (I), the user transmits and breaks the scanning loop.
		\item Else if channel $i$ is classified as (W), the user backs-off (with exponential timer) and counts down from $2^k -1$ (where $k$ is the timer window).
		\item Else if channel $i$ is classified as (J), it is added to a possible list for data transmission and the user scans the next channel.
	\end{enumerate}
	
	Note that the third step does not exist in the baseline WiFi that treats channels (W) and (J) the same way and backs off on all of these channels. At the end of scanning channels, each user performs the following:
	\begin{enumerate}
		\item If there is no (I) channel (channels are either (W) and (J)), the user selects the channel (J) with the best SINR and transmits
		\item If there is no channel (I) and (J), i.e., all channels are (W), while back-off counter is not zero, the user waits for one time slot and reduces all counters by 1.
		\item If any back-off counter is zero, the user senses that channel.
		\begin{itemize}
			\item If the channel is (I), the user transmits on that channel.
			\item Else, the user resets the back-off counter and uniformly selects a random number between $0$ and $2^{k+1}-1$.
		\end{itemize}
	\end{enumerate}
	
	\subsection{Step 5: Power Control for LPI/LPD} \label{sec:step5}
	\emph{Input}: The ID and SINR of the selected channel from Step 4.
	\\
	\emph{Output}: The transmit power given to the WiFi transmitter chain.
	\reev{Each DeepWiFi user reduces its power to the minimum to meet the minimum SINR requirement to its intended receiver for LPI/LPD capability, assuming the channel estimation from itself to its intended receiver is carried out periodically. Consider a user $i$ with the chosen transmit power $P_i$ and jammer $j$ with sensing threshold $\tau_j$, and let $h_{ij}$  denote the pathloss from user $i$ to jammer $j$. Then jammer $j$ cannot detect transmission of user $i$ if $P_i h_{ij} < \tau_j$. Note that $\tau_j$ and $h_{ij}$ are unknown to user $i$.}
	
	\subsection{Step 6: Adaptive Modulation and Coding} \label{sec:step6}
	\emph{Input}: The ID and SINR of the selected channel from Step 4.
	\\
	\emph{Output}: The MCS level and the corresponding rate for this channel given to the WiFi transmitter chain.
	
	Each user selects the modulation and coding rate for 802.11ac based on the SINR that is measured on the sensed channel. We consider the MCSs defined for the Very High Throughput (VHT) scheme of 802.11ac, as shown in Table \ref{table:MCS}.
	
	\begin{table}
		\caption{MCSs and corresponding rates for IEEE 802.11ac with one spatial stream (GI stands for guard interval).}
		\centering
		{\small
			\begin{tabular}{c|c|c|c|c}
				&&&  \multicolumn{2}{c}{\textbf{Data Rate} (Mb/s)} \\ \cline{4-5}
				\textbf{MCS}	&  \textbf{Modulation} & \textbf{Coding} & 800 ns GI & 	400 ns GI \\ \hline \hline
				0 &  BPSK & 1/2 & 6.5 & 7.2 \\ \hline
				1 &  QPSK & 1/2 & 13.0 & 14.4 \\ \hline
				2 &  QPSK & 3/4 & 19.5 & 21.7 \\ \hline
				3 &  16-QAM & 1/2 & 26.0 & 28.9 \\ \hline
				4 &  16-QAM & 3/4 & 39.0 & 43.3 \\ \hline
				5 &  64-QAM & 2/3 & 52.0 & 57.8 \\ \hline
				6 &  64-QAM & 3/4 & 58.5 & 65.0 \\ \hline
				7 &  64-QAM & 5/6 & 65.0 & 72.2 \\ \hline
				8 &  256-QAM & 3/4 & 78.0 & 86.7
				\\ \hline
			\end{tabular}
		}
		\label{table:MCS}
	\end{table}
	
\begin{figure}
		\centering
		\includegraphics[width=0.85\columnwidth]{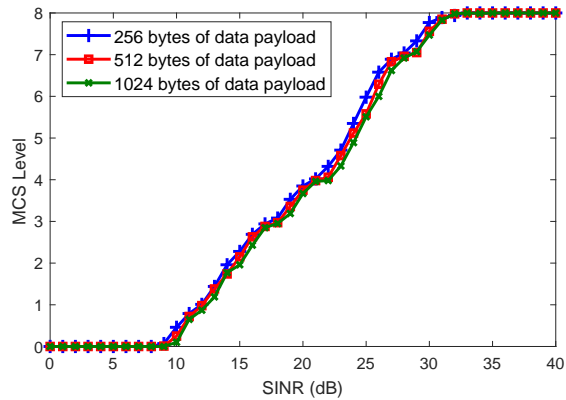}
		\caption{Best selection of MCS levels at different SINRs for different payloads.}\label{fig:MCS}
	\end{figure}

	We label the generated samples at the receiver with the MCS scheme that gives the best error rate and throughput trade-off. By design, we want lower MCS level (index) for lower SINR and higher MCS level for higher SINR. This approach provides an effective link adaptation. We determine a table that assigns the best MCS for a given SINR. This table is pre-loaded to each user that adapts its MCS according to the measured SINR. To build up this table, we initially generate the transmitted signal using a high MCS level. We transmit the signal over the channel and add noise. We estimate the channel at the receiver using the preamble and equalize the signal. The equalized samples are then used to demodulate the signal. Thus, we obtain the packet error rate by comparing the transmitted and received bits. If the packet error rate is not zero, we reduce the MCS level and repeat the same procedure. If there are still erroneous bits at the receiver for MCS 0, then we keep the MCS as 0. Fig.~\ref{fig:MCS} shows the best MCS levels (when 256, 512, and 1024 bytes of data payload are used, respectively) to maximize the rate depending on the SINR.
	
	\subsection{Step 7: Routing} \label{sec:step7}
	\emph{Input}: The set of link rates (given from Step 6) from a given user to its neighbors (note that the set of queue lengths of the user and its neighbors are obtained through information exchange in Step~7).
	\\
	\emph{Output}: The ID of the flow to serve and the ID of the neighbor selected for next-hop transmission.
	
	Each user applies the \emph{backpressure algorithm} \cite{backpressure} in a distributed setting without any centralized controller. Note that another wireless routing algorithm can be also applied here. We use the backpressure algorithm to reflect both channel and queue information in routing decisions that make use of the optimized link rates from Step~6. Each user exchanges local information on its spectrum utility with its neighbors \cite{sohraab}. Each user $i$ keeps a separate queue for each flow $s$ with backlog $Q_i^s (t)$  at time $t$. For all links, the user chooses the flow to transmit as the one with the maximum difference of queue backlogs at the receiving and transmitting ends, i.e., for each link $(i,j)$ a user $i$ chooses the flow
	\begin{align}
	s_{ij}^*= \argmax_s  [Q_i^s (t)-Q_j^s (t)]^+, 
	\end{align}
	where $[\cdot]^+ = \max⁡(\cdot,0)$. Then, we define the spectrum utility
	\begin{align}
	U_{ij} (t)=c_{ij} (t)\left[Q_i^{s_{ij}^*} (t)-Q_j^{s_{ij}^*} (t)\right]^+ 
	\end{align}
	for user $i$ transmitting to user $j$, where $c_{ij}(t)$ is the rate on link $(i,j)$ at time $t$ (depending on transmit power from Step 5 and MCS from Step 6). User $i$ transmits to the selected neighbor $j^* (t)$ that yields the maximum spectrum utility, i.e.,
	\begin{align}
	j^* (t) = \argmax_{j \in N_i}⁡ U_{ij}(t), 
	\end{align}
	where $N_i$ is the set of the next hop candidates for user $i$ to select from. Each user uses the data channel (in-band) to exchange control information with its neighbors. There is no separate control channel. Users asynchronously make decisions in a distributed setting by using the following four phases \cite{sohraab}:
	\begin{enumerate}
		\item neighborhood discovery and channel estimation,
		\item exchange of flow information updates and execution of the backpressure algorithm,
		\item transmission decision negotiation, and
		\item data transmission.
	\end{enumerate}

	\section{Simulation Setting for Network-level Performance Evaluation} \label{sec:simulation}
	There are $n = 9$ users, $m = 40$ channels, and $5$ flows generated at each simulation time. There are $n_J = 40$ jammers, each assigned to one channel. Each random jammer is independently turned on with the probability of jamming $p_J$ on every simulation time, where $p_J$ is varied between 0 to 1 with 0.05 increments. The received SINR (due to jammer and noise) is also varied with $1$~dB increments.  Each user $i$ generates traffic to a randomly selected destination $j$. The traffic rate is randomly selected from $[0,1]$ Mb/s  with average rate $r_{ij} =500$ kbps for a source-destination pair $(i,j)$. The simulator is implemented in MATLAB. Simulation time is 100 seconds. The baseline WiFi features are implemented using the MATLAB WLAN System Toolbox. The deep learning code is implemented in TensorFlow.

	We compare the performance of DeepWiFi with the baseline WiFi (without deep learning or jamming resistance) and jamming-resistant MAC protocol JADE \cite{Jade}. JADE is asymptotically optimal in the presence of adaptive adversarial jamming. As the baseline WiFi is not designed to mitigate any jamming, its comparison with DeepWiFi quantifies the effect of jamming defense. In DeepWiFi, we use the exponential backoff defined in the IEEE 802.11ac standard. The comparison of DeepWiFi with JADE quantifies the difference in the channel access and different solutions spaces. Channel access in JADE does not differentiate between a WiFi or a jamming signal, whereas DeepWiFi does. Therefore, the solution spaces are different. We show that DeepWiFi achieves major performance improvement over the baseline WiFi and JADE. In all three algorithms, we use adaptive modulation and coding and backpressure routing algorithm for a fair comparison. 
	
\begin{figure}
		\centering
		\includegraphics[width=\columnwidth]{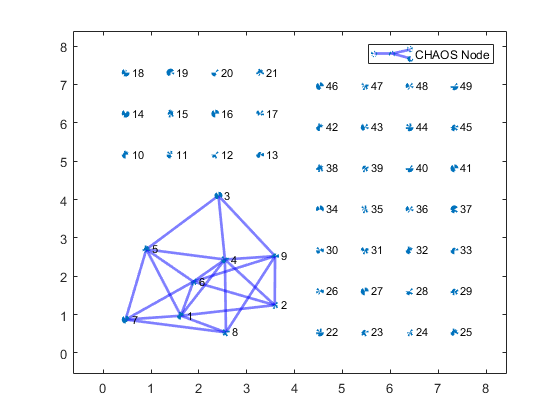}
		\caption{Network topology for simulations.}\label{fig:topology}
	\end{figure}

	The network topology is shown in Figure~\ref{fig:topology} where we deployed nine users uniformly at random over a given area. Since friendly users do not interfere with each other due to the backoff mechanism and signal classification, their locations do not affect the system performance. The friendly users (DeepWiFi or baseline WiFi) are labeled with IDs~1-9. Each link represents the set of neighbors (depending on the SINR threshold less than $0$~dB).
	The 40 jammers are depicted as users without links on the right and top of Figure~\ref{fig:topology} (with user ID 10-49). Each jammer is responsible to jam a single channel (out of 40 channels). For instance, jammers with ID 10, 11, 12 would respectively jam  channels 1, 2, 3 and so on. When the jammer is on (off), it is depicted as a red (blue) user.
	
	Consider the following metrics to measure the performance:
	\begin{enumerate}
		\item \emph{Average throughput} (Mb/s) per user: The average throughput rate that each user can achieve during the simulation.
		\item \emph{Cumulative throughput}: The average throughput of the network for all users during the simulation.
	\end{enumerate}

	\section{Performance Results} \label{sec:performance}
	\subsection{Probabilistic (random) jammer}
\begin{figure}
		\centering
		\includegraphics[width=0.8\columnwidth]{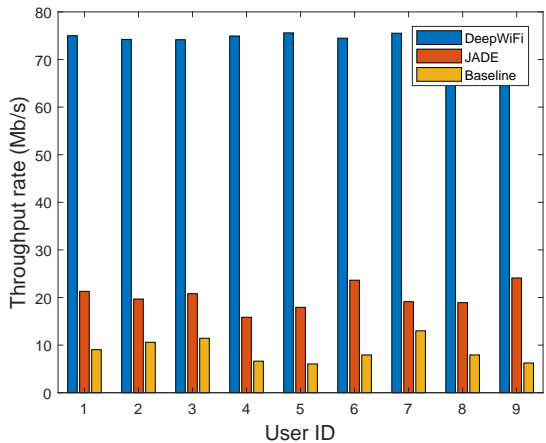}
		\caption{Throughput of individual users (when $p_J = 0.7$) for fixed SINR ($0$~dB).}\label{fig:sim3}
	\end{figure}

\begin{figure}
		\centering
		\includegraphics[width=0.875\columnwidth]{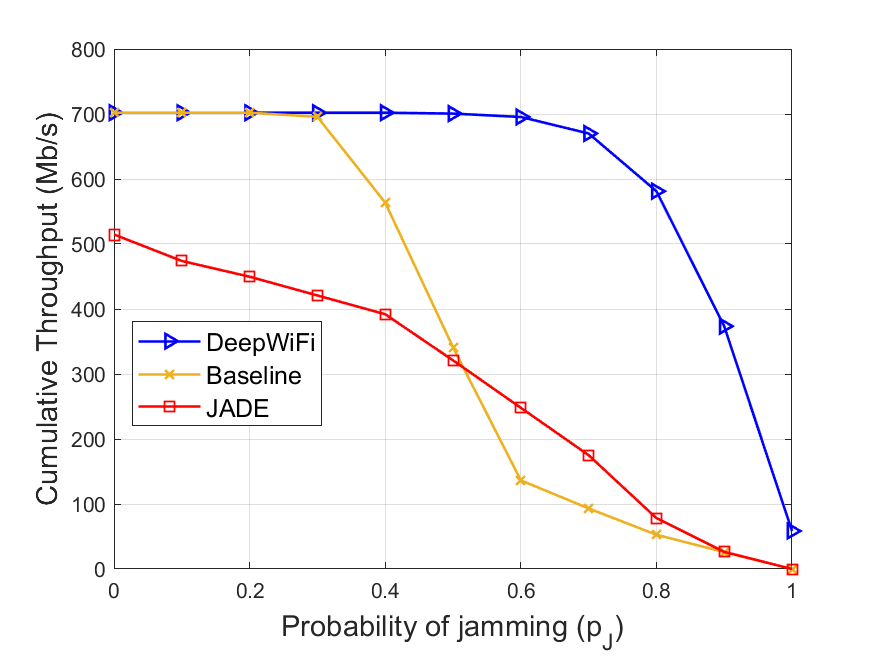}
		\caption{Cumulative throughput as a function of $p_J$ for fixed SINR ($0$~dB).}\label{fig:sim1}
	\end{figure}

\begin{figure}
		\centering
		\includegraphics[width=0.875\columnwidth]{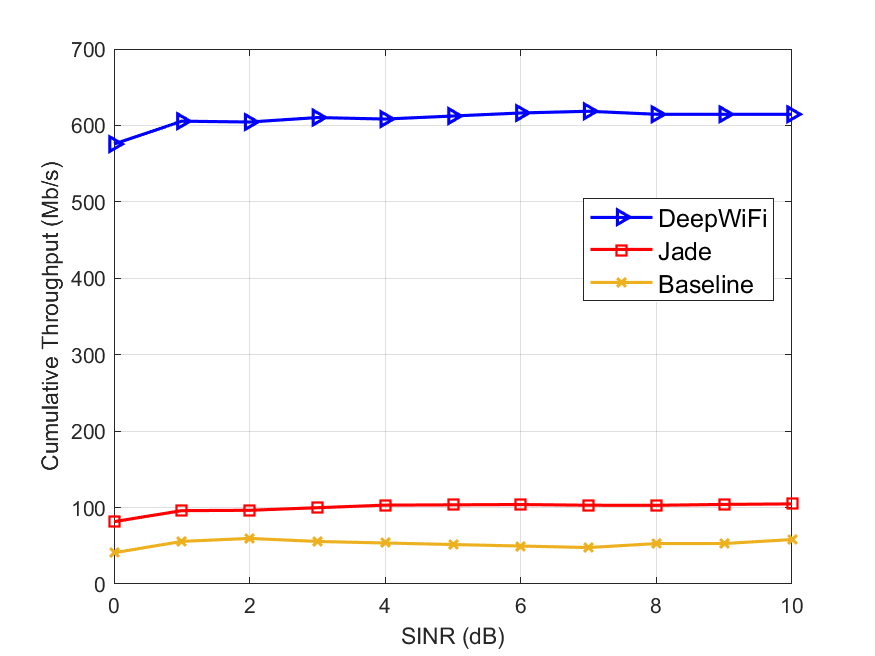}
		\caption{Cumulative throughput as a function of SINR for $p_j = 0.8$.}\label{fig:sim2}
	\end{figure}

	First, we fix the SINR to $0$~dB and start with random jammers. The throughput of individual users (when $p_J = 0.7$) for fixed SINR ($0$~dB) is shown in Figure~\ref{fig:sim3}. \rav{The end-to-end network throughput is shown in Figure~\ref{fig:sim1} as a function of $p_J$}. For small $p_J$, both DeepWiFi and baseline WiFi achieve the same throughput, since all users can find idle channels without internal or external interference. JADE is run with its default parameters  according to \cite{Jade}. Due to its small channel access probability initialization, JADE starts off worse than the baseline WiFi and DeepWiFi. As $p_J$ increases beyond $0.3$, the throughput of baseline WiFi starts dropping sharply, while DeepWiFi sustains its throughput and provides major throughput gains relative to baseline WiFi. For $p_J \geq 0.5$, JADE resists to jamming better than the baseline WiFi by adjusting the channel access probability and outperforms it while performing worse than DeepWiFi. Even when all channels are jammed all the time (with $p_J =1$), DeepWiFi can achieve the network end-to-end throughput close to $70$ Mbs, while the throughputs of baseline WiFi and JADE are zero (note that fluctuations are due to randomness in channels and interference effects). DeepWiFi activates more links than baseline WiFi and JADE, as users back off on the jammed channels in both schemes, whereas DeepWiFi allows users to transmit on the jammed channels (if no idle channel available) in the degraded mode. We start changing the SINR due to jammer and noise effects. The cumulative throughput is shown in Fig.~\ref{fig:sim2} as a function of SINR at $p_J=0.8$. The cumulative throughput slowly increases with SINR for all schemes and DeepWiFi outperforms others over all SINRs. Also, note that the cumulative throughput term in this paper accounts for multiple hops across the network and represents the end-to-end network throughput.
	
	\subsection{Static sensing-based jammer}
	\label{sec:static_jammer}
	So far, we considered random (probabilistic) jammers that are turned on with some fixed probability. Next, we evaluate the effect of sensing-based jammers and the performance of power control for LPI/LPD.  Here, we consider a static jammer that has a constant sensing threshold $\tau$, whereas an adaptive jammer (discussed in Section~\ref{sec:adaptive_jammer} can dynamically adjust its sensing threshold. Let $r_k^{(n)}$ denote the jammer's received signal on channel $n$ at time $k$.
		The static sensing-based jammer turns on if it detects a signal on the channel greater than or equal to a threshold $\tau$, that is $r_i^{(n)} \geq \tau$, else it is turned on with a fixed probability of jamming, $p_J$.

\begin{figure}
		\centering
		\includegraphics[width=0.8\columnwidth]{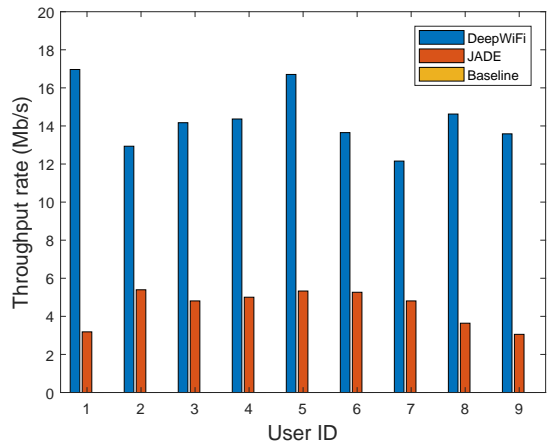}
		\caption{Individual throughput (Mb/s) of DeepWiFi and baseline WiFi when sensing-based jammers are on, $p_{J} = 0.7$ and $\tau_J =10$~dB.}\label{fig:no1}
	\end{figure}

\begin{figure}
		\centering
		\includegraphics[width=0.825\columnwidth]{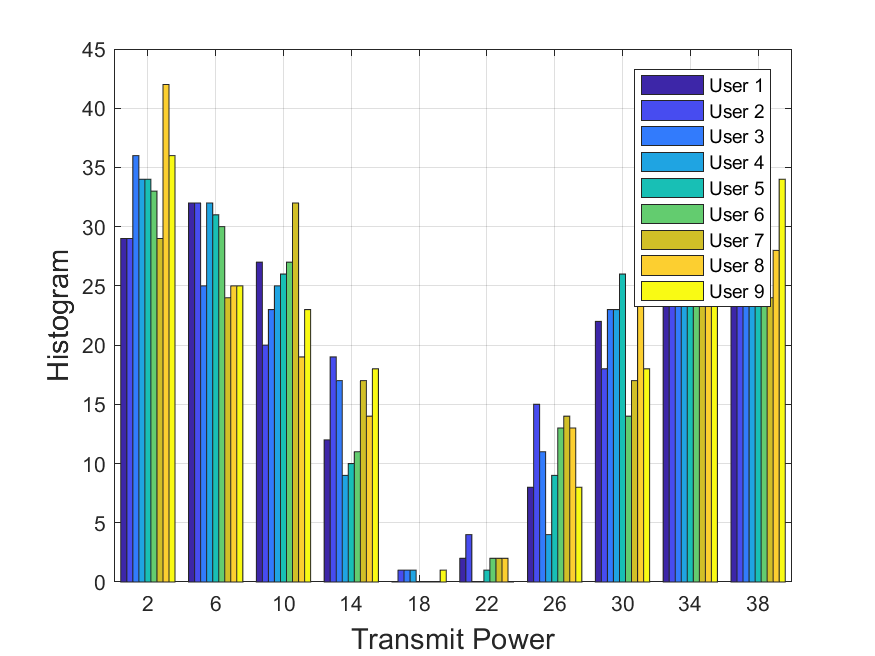}
		\caption{The transmit power is adjusted for LPI/LPD.}\label{fig:yes1}
	\end{figure}

\begin{figure}
		\centering
		\includegraphics[width=0.75\columnwidth]{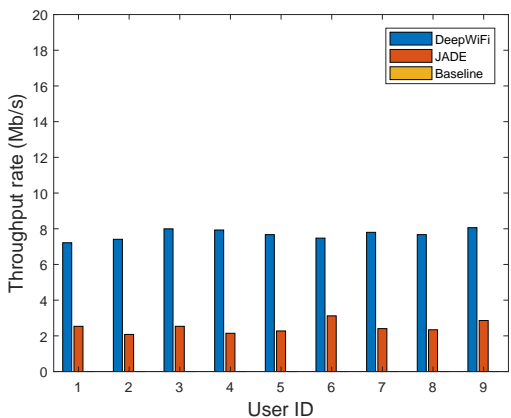}
		\caption{Individual throughput (Mb/s) of DeepWiFi and baseline WiFi when sensing-based jammers are on, $p_J = 0.7$, SINR is $5$ dB, $\tau_J =2$~dB, and the transmit power is adjusted for LPI/LPD.}\label{fig:yes2}
	\end{figure}

	In our experiments, the power detection threshold of sensing-based jammer is varied from $2$~dB to $10$~dB with $1$~dB increments. 
	Fig.~\ref{fig:no1} shows the average individual throughput (Mb/s) of DeepWiFi, JADE, and baseline when sensing-based jammers are on, $p_J = 0.7$ and $\tau_J =5$~dB. We observe that the sensing-based jammer reduces the throughput of baseline WiFi to zero as  users end up backing off indefinitely.
	Next, we present the results for transmit power control to DeepWiFi to demonstrate the LPI/LPD capability. DeepWiFi with transmit power control for LPI/LPD can avoid jammers by operating below $\tau_J$. Fig.~\ref{fig:yes1} shows the histogram of transmit power per user. We observe that about half of the transmissions are with low power to provide LPI/LPD capability, whereas the other half are with high power to support communication over the jammed channels. As a result, DeepWiFi with LPI/LPD can achieve higher rates higher compared to baseline WiFi, JADE, and DeepWiFi without LPI/LPD. Fig.~\ref{fig:yes2} shows the average individual throughput (Mb/s) of DeepWiFi and baseline WiFi when sensing-based jammers are on, $p_J = 0.7$, $\tau_J = 2$~dB, and the transmit power is adjusted for LPI/LPD. We observe that as the jammer becomes more reactive, i.e., the detection threshold of jammer $\tau_J$ decreases from 10~dB to 2~dB, the throughput of DeepWiFi and JADE decrease while the baseline WiFi's throughput diminishes completely.

	\begin{figure}
		\centering
		\includegraphics[width=0.8\columnwidth]{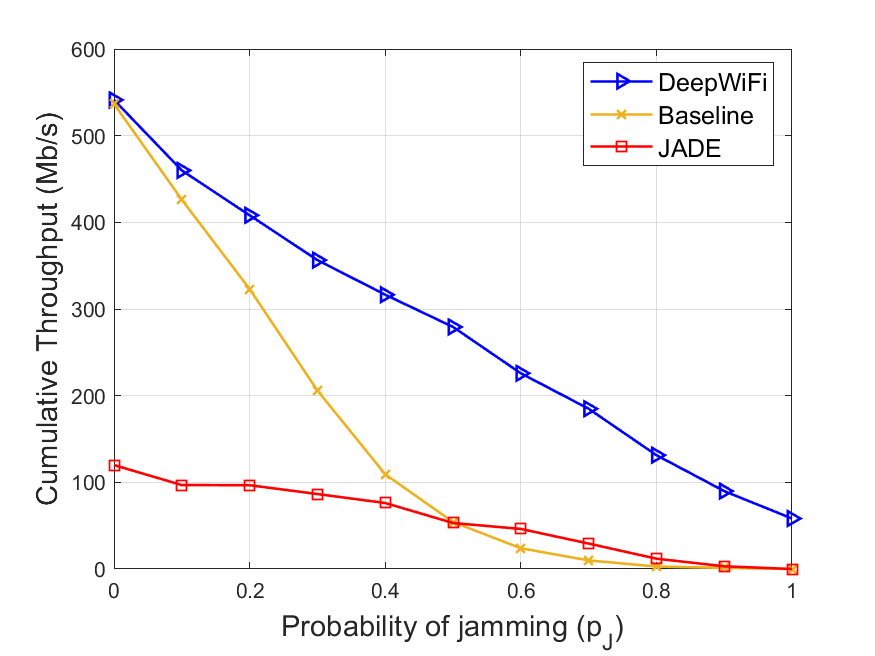}
		\caption{Cumulative throughput (Mb/s) as a function of $p_J$ when the number of channels is less than the number of users in the presence of static sensing-based jammers.}\label{fig:sum_throughput_kn}
	\end{figure}
	
We also evaluate DeepWiFi when there are more users than available channels. We consider 9 users and 9 traffic flows sharing 6 channels. Figure~\ref{fig:sum_throughput_kn} presents the cumulative throughput as a function of jamming probability $p_J$ for DeepWiFi, baseline WiFi, and JADE. We observe a linear decrease in cumulative throughput as the jamming probability increases. Note that the cumulative throughput of DeepWiFi depends on the jamming probability, the number of users, number of channels, backoff mechanism, jamming power, and signal classification accuracy. Comparing Fig.~\ref{fig:sim1} and Fig.~\ref{fig:sum_throughput_kn}, we observe that the slope of cumulative throughput versus jamming probability shows logarithmic-like behavior when there are fewer users than channels, whereas the slope is almost linear when there are more users than channels (i.e., effect of congestion increases).
	
	\subsection{Adaptive jammer}\label{sec:adaptive_jammer}
	Next, we consider the adaptive jammer that dynamically adjusts the sensing threshold $\tau_t$ at time $t$ by observing the channel access patterns of WiFi users.
	The channel utilization on channel $n$ (as the jammer perceives it) at time $k$ is  $\mathds{1}(r_k^{(n)} \geq \tau_k)$, where $\mathds{1}(x)$ denotes the indicator function, that is, 1 when $x$ is True and 0 otherwise.
	At time $t$, the adaptive jammer's utility function that includes the channel utilization and its own power consumption is defined as
	\begin{align}
	g(\textbf{r}(t),\textbf{p}(t)) = \frac{1}{T_0} \sum_{k=t-T_0+1}^{t} \sum_{n=1}^N \left( \mathds{1}(r_k^{(n)} \geq \tau_k) + w \cdot  p_k \right),
	\end{align}
	where $p_k$ is the jamming power at time $k$ and $w \geq 0$ is a weighting constant to balance the tradeoff between jammer power consumption and WiFi channel utilization. The vectors $\textbf{r}(t)$ and $\textbf{p}(t)$ denote the received signal and transmit power values of the jammer, respectively, over a time window of $T_0$ time instances ending at time $t$ and $N$ channels, namely $\textbf{r}(t) = \{r_k^{(n)} , 1 \leq n \leq N, t-T_0+1 \leq k \leq t\}$ and  $\textbf{p}(t) = \{p_k, t-T_0+1 \leq k \leq t\}$. At time $t$, the jammer transmits when its received signal on channel $n$ is greater than or equal to its sensing threshold, namely $r_t^{(n)} \geq \tau_t$. Using a sufficiently small step size constant $\Delta > 0$, the adaptive jammer updates its sensing threshold as $\tau_{t+1} = \tau_t + \frac{\Delta}{t}$ if $g(\textbf{r}(t-1),\textbf{p}(t-1)) > g(\textbf{r}(t),\textbf{p}(t))$, else $\tau_{t+1} = \left[ \tau_t - \frac{\Delta}{t} \right]^+$.

\begin{figure}[t!]
		\centering
		\includegraphics[width=0.825\columnwidth]{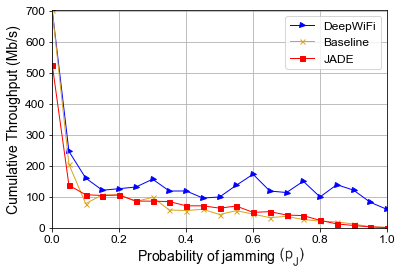}
		\caption{\reev{Cumulative throughput (Mb/s) as a function of $p_J$ in the presence of adaptive jammers.}}\label{fig:adaptive_jammer1}
	\end{figure}

	\reev{Fig.~\ref{fig:adaptive_jammer1} shows the performance of DeepWiFi, baseline, and JADE under adaptive jamming. We took the initial sensing threshold as $\tau_0=1$, weighting constant as $w=1$, and step size constant $\Delta = 0.5$. We observe that the cumulative throughput reduces significantly in the presence of adaptive jammers in all cases, while DeepWiFi still provides the best performance compared to the baseline WiFi and JADE.}
	
	\reev{Overall, the results  evaluated in the presence of three different types of jammers indicate that DeepWiFi provides reliable and robust communication.}

	\section{Implementation, Complexity, and Overhead Aspects}\label{sec:implementation}
	\rev{DeepWiFi can be implemented in the  kernel or in the WiFi card. Open-source firmwares such as Nexmon} \cite{nexmon} \rev{can extract the I/Q data out of the WiFi such that the I/Q data can be processed in the kernel. System-on-chip (SOC) solutions built for IoT such as Qualcomm QCS603 / QCS605} \cite{QualcommSoC} \rev{can be also used to extract the I/Q data that can be processed in the kernel (or can be moved to an additional FPGA or ARM for further processing). These SOCs include 802.11ac WiFi and a neural processing engine (for deep learning) integrated in the chip that can be used for RF communication and deep learning. The training is usually performed offline and the trained models are ported to the kernel or to the other platforms for inference.}
	
	\rev{DeepWiFi can get the training data in two possible ways. First,  channels and signals can be generated by MATLAB WLAN Toolbox (as discussed in Section}~\ref{sec:channel}\rev{), and they can be used to build the training data. Second, a real 802.11.ac WiFi card can be used to make over-the-air transmissions that are captured by an SDR, and the channels and signals can be used for training.}

	\reev{To evaluate the time complexity and overhead, we implemented each deep learning task in a low-cost embedded system, NVIDIA Jetson Nano Developer Kit \cite{NvidiaNano}}.
	\reev{For an efficient deployment, we converted the Tensorflow code to TensorRT \cite{TensorRT}, inference optimizer for NVIDIA's embedded systems. We repeated the tests 1000~times to calculate the average inference time.} \reev{For example, front end processing of each data sample takes $0.09$~msec}. A typical frame in the 802.11ac standard is $5.484$~msec. \cite{Gast}. \reev{Hence, the front end processing time is measured as small as $1.6\%$ of the 802.11 frame.}
	\rav{The reported processing time is for the case when there is already data available to be processed. As the performance of the off-the-shelf hardware can be degraded when there are other processes (such as other WiFi operations) running in the background, we used the $\texttt{stress-ng}$ software\cite{streesng}  to stress test the memory use of the embedded GPU by generating background processes emulating high load conditions.
		We observed that the processing time for deep learning operations increased 33\% but remained small relative to MAC frame length. Note that the overhead and processing times on embedded platforms can be further reduced by customized ASIC chips and FPGAs.}
	
	\reev{For the signal classification task, we observed that the FNN model takes $0.009$~msec on average to predict each sample point. On the other hand, the CNN model takes $0.035$~msec to predict a sample over 1000~repetitions.} As a comparison between the FNN and CNN models, both architectures have similar performance. Since the CNN architecture has more layers, it takes approximately four times longer to process a single waveform in inference time. CNNs in return have a smaller footprint on the hardware. 
	Note that the difference in hardware footprint is expected since the FNNs use fully connected layers to reduce dimensions from one layer to another, whereas the CNNs employ convolutional layers that utilize feature maps in a sliding window manner. For more discussion on the number of parameters in FNN and CNN models, we refer the reader to \cite[p.~357]{Geron}. 
	
	\rev{For the signal authentication task, the MCD outlier detection takes $0.0556$~msec, one-class SVM takes $52.813$~msec, and isolation forest takes $33.521$~msec.}
	
	\reev{Adding these components, we observe that the overall processing overhead for the front end processing, signal classification using FNN, and signal authentication using MCD is $0.1546$~msec, which is only $2.8\%$ of the 802.11ac frame.}
	
	For the routing overhead, the backpressure algorithm exchanges typical messages as in the regular distance-vector routing algorithms. There is no closed form expression for the number of message exchanges and the exact number depends on the network and traffic conditions. The backpressure algorithm was implemented with SDRs in \cite{Soltani15} that empirically evaluated the number of message exchanges.
	
	\section{Conclusion} \label{sec:conclusion}
	We presented the DeepWiFi protocol that applies machine learning, and in particular deep learning to adapt WiFi to spectrum dynamics and provides major throughput gains compared to baseline WiFi \rev{and another jamming-resistant MAC protocol}. DeepWiFi is designed to mitigate out-of-network interference effects from probabilistic and sensing-based jammers. Built upon the PHY transceiver chain of IEEE 802.11ac, DeepWiFi  provides the decision parameters to WiFi transceiver without changing the PHY and the MAC frame format. DeepWiFi applies deep learning-based autoencoder to extract features and applies deep neural networks (FNN or CNN) to classify signals as WiFi, jammed, or idle. The signals classified as WiFi are further processed through RF fingerprinting that identifies hardware-based features by machine learning and authenticates legitimate WiFi signals. Using these signal labels on sensed channels, DeepWiFi supports users to use idle channels, back-off on channels used by legitimate WiFi signals, and access (if needed in the degraded mode) channels that are occupied by signals other than legitimated WiFis. DeepWiFi users optimize their transmit powers for LPI/LPD and their MCS to maximize link rates, and their routing decisions by backpressure algorithm. We simulated DeepWiFi in a distributed network using the channels and signals generated by MATLAB WLAN Toolbox. We showed that DeepWiFi helps WiFi users sustain their throughput with major performance gain relative to baseline WiFi \rev{and another jamming-resistant MAC protocol} especially when channels are likely to be jammed and the SINR is low.
	
\section*{Appendix} 

\subsection{Other Dimensionality Reduction Techniques}\label{Section:DimReduction}
\begin{figure}[ht!]
	\centering
	\includegraphics[width=0.8\columnwidth]{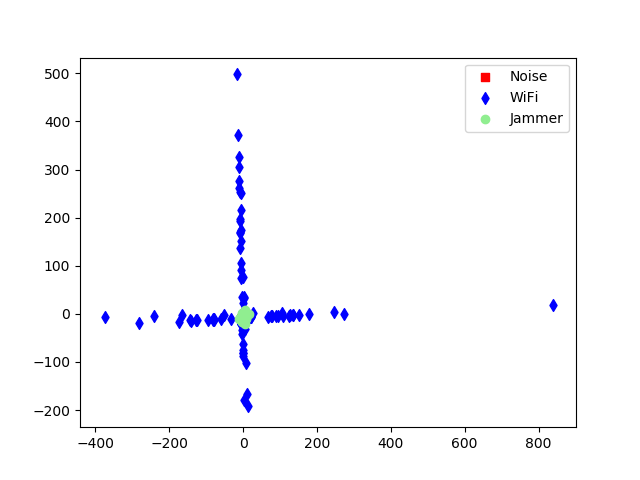}
	\caption{Dimensionality reduction to 2 features using PCA.}\label{fig:pcaN2}
\end{figure}

\begin{figure}[ht!]
	\centering
	\includegraphics[width=0.8\columnwidth]{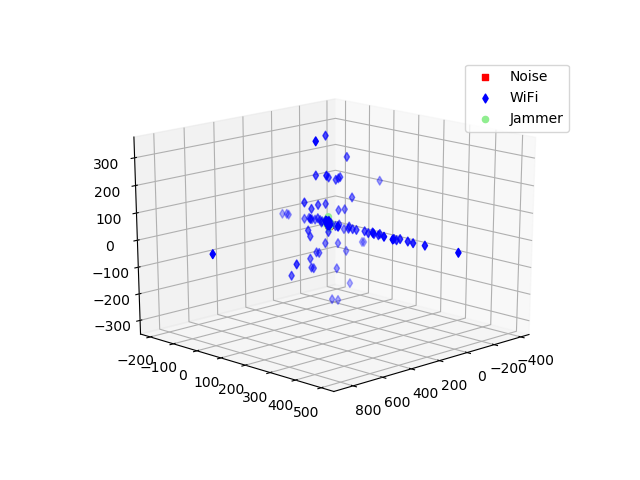}
	\caption{Dimensionality reduction to 3 features using PCA. All signal types are plotted in the same figure.}\label{fig:pcaN3all}
\end{figure}

\begin{figure}[ht!]
	\centering
	\includegraphics[width=0.95\columnwidth]{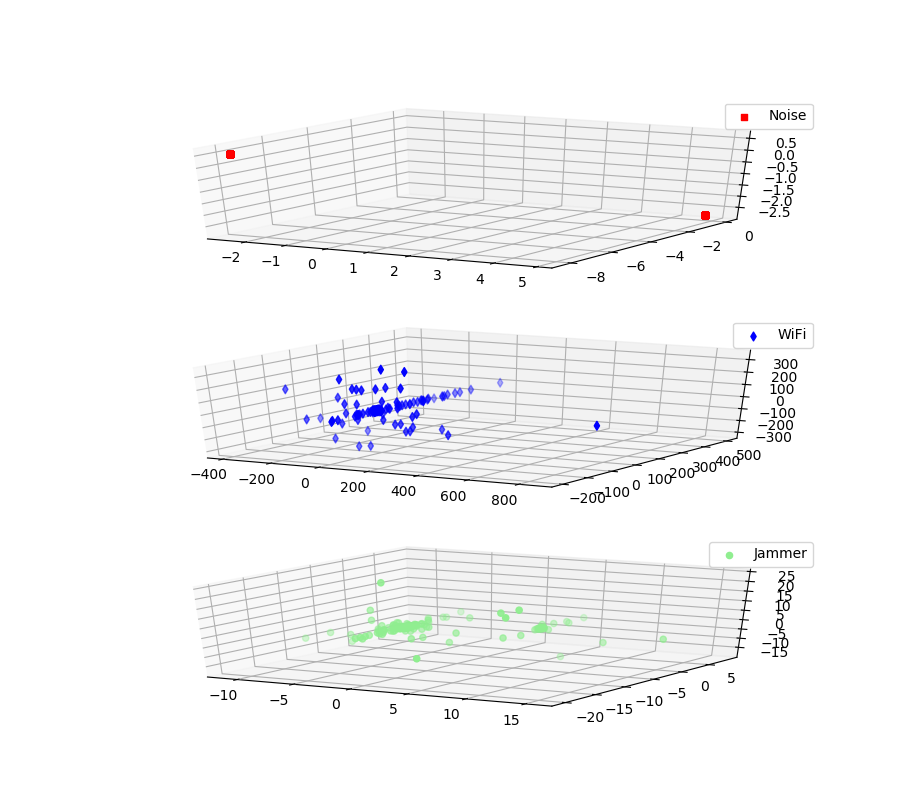}
	\caption{Dimensionality reduction to 3 features using PCA. Three signal types are individually plotted.}\label{fig:pcaN3sub}
\end{figure}

In addition to an autoencoder, we have also tested the separability of the noise, WiFi, and jamming signals using other dimensionality reduction methods such as Principal Component Analysis (PCA) \cite{PCA} and t-distributed stochastic neighbor embedding (tSNE) \cite{tSNE}. We conclude that these techniques cannot separate signal types of interest. The PCA method is an orthogonal linear transformation that converts the data into a new coordinate system that is defined by the principal coordinates \cite{PCA}. The first component of the PCA can be found by solving
\begin{align}
\bm{w}_1 = \arg \max \frac{\bm{w}^T \bm{X}^T \bm{X} \bm{w}}{\bm{w}^T \bm{w}}. \label{eqn:PCA}
\end{align}
This term is also called as the Rayleigh quotient and the maximum is attend at the largest eigenvalue of the matrix $\bm{X}^T \bm{X}$. The other components of the PCA are found using a similar procedure but by subtracting the effects of the previous components. For example, to calculate the $k$th PCA component, $k-1$ PCA components are subtracted from $X$ before solving (\ref{eqn:PCA}). Thus, we first perform the subtraction
\begin{align}
\bm{\hat{X}}_k = \bm{X} - \sum_{j=1}^{k-1} \bm{X} \bm{w}_j \bm{w}_j^T,
\end{align}
and then solve
\begin{align}
\bm{w}_k = \arg \max \frac{\bm{w}^T \bm{\hat{X}}^T \bm{\hat{X}} \bm{w}}{\bm{w}^T \bm{w}}.
\end{align}
First, we use PCA to reduce the dimensions to two, $N=2$. Figure~\ref{fig:pcaN2} illustrates the distributions of three signal types. We observe that WiFi signal is distributed along the two eigenvector directions, whereas the jammer and noise is clustered around the origin. Next, we increase the dimension size to three, $N=3$ and try to see if they are separated in 3-dimensions. Figures~\ref{fig:pcaN3all} and~\ref{fig:pcaN3sub} illustrate the distributions for three components $N=3$. Figure~\ref{fig:pcaN3all} demonstrates all three types of points on top of each other whereas Figure~\ref{fig:pcaN3sub} plots each type in a different plot. We observe that while the WiFi signal is well distributed in three dimensions, the jamming and noise signals are clustered in smaller values (in a horizontal plane around $x \in [-10,15]$ and $y \in [-20,5]$) which makes it hard to distinguish. We conclude that WiFi and jamming signals are not linearly separable and PCA is not a good candidate to differentiate the clusters for these three signal types.


\begin{figure}[t!]
	\centering
	\includegraphics[width=0.8\columnwidth]{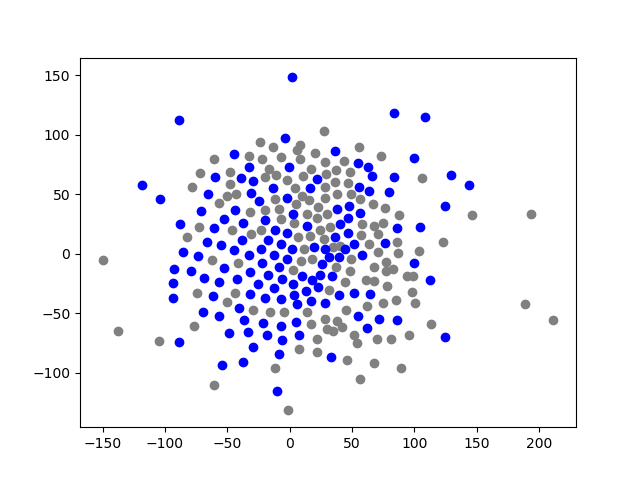}
	\caption{Dimensionality reduction  to 2 features using t-SNE.}\label{fig:tSNE2}
\end{figure}
\begin{figure}[t!]
	\centering
	\includegraphics[width=0.8\columnwidth]{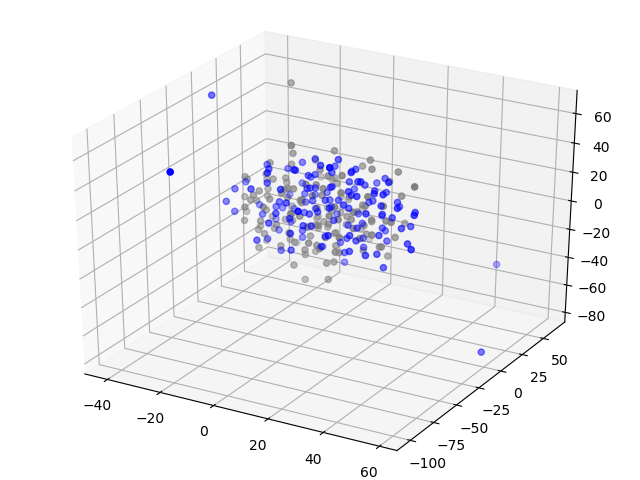}
	\caption{Dimensionality reduction to 3 features using t-SNE.}\label{fig:tSNE3}
\end{figure}

Next, we try another commonly used dimensionality reduction technique called t-SNE. Unlike PCA, t-SNE uses the joint probabilities between data points and tries to minimize the Kullback-Leibler (KL) divergence between the joint probabilities of the low-dimensional embedding and the high-dimensional data \cite{tSNE}. The t-SNE method first computes conditional probabilities $p_{j|i}$ between the input data
\begin{align}
p_{j|i} = \frac{\exp(-||\bm{x}_i - \bm{x}_j||^2/2\sigma_i^2)}{\sum_{k\neq i} \exp(-||\bm{x}_i - \bm{x}_k ||^2 /2\sigma_i^2)}.
\end{align}
Then,
\begin{align}
p_{ij} = \frac{p_{j|i} + p_{i|j}}{2N},
\end{align}
where $N$ denotes the number of sample points. Let $\bm{z}_1, ..., \bm{z}_N$ denote the representations of the input dataset in the reduced dimensions such that $\bm{z}_i \in \mathbb{R}^d$ where $d$ denotes the dimensions to be reduced to. In this case, we evaluate for $d=2$ and $d=3$. The similarity of the representations $\bm{z}_i$ and $\bm{z}_j$ in $d$-dimensions is
\begin{align}
q_{ij} = \frac{(1 + ||\bm{z}_i - \bm{z}_j ||^2)^{-1}}{\sum_{k\neq i} (1 + ||\bm{z}_i - \bm{z}_k ||^2)^{-1}}.
\end{align}
Using these similarity measures, the t-SNE method employs the KL divergence of the reduced dimension distribution $Q$ from the data distribution $P$ and solves
\begin{align}
KL(P||Q) = \sum_{i\neq j} p_{ij} \log\left(\frac{p_{ij}}{q_{ij}}\right).
\end{align}
The KL divergence is used as a cost function in t-SNE.

We apply the t-SNE method to reduce the dimensionality of WiFi and jamming signals, while the noise is kept aside for this study for visual clarity. While t-SNE is particularly well suited for visualization of high-dimensional dataset, Figures~\ref{fig:tSNE2} and \ref{fig:tSNE3} demonstrate that WiFi and jamming signals are not separable with a smoothline between clusters. Thus, we do not use the t-SNE.

\subsection{Signal Classification}
\begin{figure}
	\centering
	\begin{subfigure}[b]{\columnwidth}
		\includegraphics[width=\columnwidth]{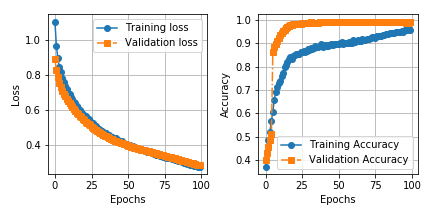}
		\caption{FNN}
	\end{subfigure}
	\begin{subfigure}[b]{\columnwidth}
		\includegraphics[width=\columnwidth]{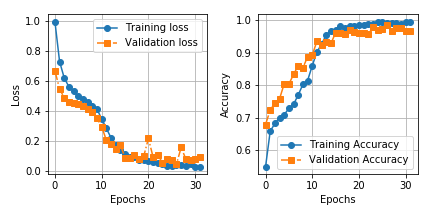}
		\caption{CNN}
		\label{fig:CNN}
	\end{subfigure}	
	\caption{\rev{Cross-entropy loss and accuracy of both architectures for signal classification.}}
	\label{fig:FNN_CNN}
\end{figure}

\rev{As discussed in Section~5.3, the generated dataset is divided into $80\%$ training and $20\%$ testing. We provide the accuracy and loss functions of both FNN and CNN architectures for the training and test sets in Figures}~\ref{fig:FNN_CNN}\rev{(a)-(b). The results in these figures validate that the final model does not suffer from overfitting or underfitting.}

\section*{Acknowledgment}
We would like to acknowledge SBIR data rights.\footnote{Contract No.: W91CRB-17-P-0068
	Contractor Name: Intelligent Automation, Inc.
	Contractor Address: 15400 Calhoun Drive, Suite 190, Rockville, MD 20855
	Expiration of SBIR Data Rights Period: 02/23/2023, unless extended by additional funding, in which case the SBIR Data Rights Period will expire 5 years following termination of such additional funding. The Government's rights to use, modify, reproduce, release, perform, display, or disclose technical data or computer software marked with this legend are restricted during the period shown as provided in paragraph (b)(4) of the Rights in Noncommercial Technical Data and Computer Software--Small Business Innovative Research (SBIR) Program clause contained in the above identified contract. No restrictions apply after the expiration date shown above. Any reproduction of technical data, computer software, or portions thereof marked with this legend must also reproduce the markings.}

\bibliographystyle{IEEEtran}
\bibliography{IEEEabrv,references2}	
\end{document}